\documentclass[10pt,reqno,a4paper]{amsart}
\usepackage[cp1251]{inputenc}
\usepackage[russian]{babel}
\usepackage{amssymb,graphics}
\newcommand{\mfrac}[2]
{\mbox{\footnotesize$\displaystyle\frac{\raisebox{-0.1em}{\mbox{$#1$}}}{
\raisebox{0.1em}{\mbox{$#2$}}}$}}

\newcommand{\reff}[1]{\mbox{$(\ref{#1})$}}

\newcommand{\gk}{\gamma_k^{}}
\newcommand{\pot}{{\mbox{\footnotesize$\raisebox{0.07em}{\mbox{\rm [}}
U\raisebox{0.07em}{\mbox{\rm ]}}$}}}

\newcommand{\potxt}{{\rm\mbox{\footnotesize$[U(x,t)]$}}}
\newcommand{\sss}{\scriptscriptstyle}
\newcommand{\ds}{\displaystyle}

\newcommand{\thetaAB}[2]
{\theta \raisebox{0.04em}{\mbox{$ \scalebox{0.5}[0.9]{\big[}
\!\!\raisebox{0.1em} {\scalebox{0.8}{\mbox{\tiny$
\begin{array}{c}#1\\#2\end{array}$}}}
\!\! \scalebox{0.5}[0.9]{\big]}$}}}

\newcommand{\varthetaAB}[2]
{\vartheta \raisebox{0.04em}{\mbox{$ \scalebox{0.5}[0.9]{\big[}
\!\!\raisebox{0.1em} {\scalebox{0.8}{\mbox{\tiny$
\begin{array}{c}#1\\#2\end{array}$}}}
\!\! \scalebox{0.5}[0.9]{\big]}$}}}

\begin{document}

\title{Об интегрируемости конечнозонных  потенциалов}

\author{\sc  Ю.\,Брежнев}
\email{brezhnev@mail.ru}

\begin{abstract}
Для спектральных задач, определяемых обык\-но\-вен\-ны\-ми
дифференциальными уравнениями, мы формулируем конечнозонные
потенциалы как точно решаемые в квадратурах по Пикару--Вессио без
привлечения специальных функций и предполагаем, что этот класс
единственный. Идеология восходит к работам Драша 1919 г., которая
однако не нашла отражения в современной литературе. Приводя новые
примеры, мы демонстрируем технику получения не\-об\-хо\-ди\-мых
сос\-тав\-ля\-ю\-щих точной интегрируемости: новой формулы для
$\Psi$-функции, главной формулы следов и задачи обращения абелевых
интегралов.\end{abstract}

\maketitle

\tableofcontents

\section{Введение}
\noindent Теория конечнозонного (алгебро-геометрического)
интегрирования нелинейных дифференциальных уравнений в частных
про\-из\-вод\-ных, динамических систем и спектральных задач в
современном понимании появилась 30 лет назад в работах С.\,Новикова
\cite{19,26}, В.\,Матвеева \cite{matv, 4,21}, А.\,Итса
\cite{57,17,21},  Б.\,Дубровина \cite{7,8,70}, И.\,Кричевера
\cite{24,25}, П.\,Лакса \cite{55}, Х.\,МакКина \cite{37} и др. как
периодическое обобщение метода обратного спектрального
преобразования ({\sc мозр}) для убывающих потенциалов \cite{39}. В
первых же работах по этой тематике \cite{19,26} стало ясно, что
аналогами и обобщениями солитонов являются гладкие вещественные
потенциалы в спектральной задаче уравнения Шрёдингера с непрерывным
спектром, образованным конечным числом запрещенных зон --- областей
спектрального параметра $\lambda$, где $\Psi$-функция, как функция
от $x$, неограниченно растет. Дальнейшее развитие   \cite{29,39}
привело к тому, что конечнозонными, уже произвольных матричных и
скалярных спектральных задач, стали называть потенциалы, для которых
спектральный параметр является мероморфной функцией
$\lambda(\mathcal{P})$ на римановой поверхности алгебраической
кривой конечного рода $W(\mu,\,\lambda)=0$, а $\Psi$-функция, как
функция $\Psi(\mathcal{P})$, является функцией экспоненциального
типа \cite{15,32} имеющей одну или несколько существенных
особенностей заданного вида \cite{4,24}. Этот результат Итса
\cite{57,56} и Кричевера \cite{24}, известный сейчас как аксиоматика
функции Бейкера--Ахиезера (БА), означает в частности, что все
скалярные спектральные задачи вида
\begin{equation}\label{gd}
\widehat L\Psi\equiv\frac{d^n}{dx^n}\,\Psi+u_2^{}(x)\,
\frac{d^{n-2}}{dx^{n-2}}\,\Psi+\cdots+ u_n(x)\,\Psi=\lambda\,\Psi\,,
\end{equation}
в классе конечнозонных потенциалов, имеют универсальное
нормированное решение
\begin{equation}\label{theta}
\Psi\big(x;\,\lambda(\mathcal{P})\big)=
\frac{\Theta(\boldsymbol{D})\, \Theta\big(x\,\boldsymbol{U}+
\boldsymbol{D}+\boldsymbol{A}(\mathcal{P})\big )}
{\Theta\big(\boldsymbol{A}(\mathcal{P})+ \boldsymbol{D}
\big)\,\Theta\big(x\,\boldsymbol{U}+\boldsymbol{D}\big)}\,
e^{\Omega(\mathcal{P})x}\,,
\end{equation}
а отличия их друг от друга  содержатся в строении соответствующих
алгебраических кривых. Здесь,
$\Theta(\boldsymbol{z})\equiv\Theta(\boldsymbol{z}|B)$
--- тэта-ряд, построенный по $B$-матрице периодов кривой
$W(\mu,\lambda)=0$, $\Omega(\mathcal{P})$
--- нормированный абелев интеграл 2-го рода с единственным
полюсом 1-го порядка, $\boldsymbol{U}$ --- его
$\boldsymbol{b}$-периоды, $\mathcal{P}$ --- точка на кривой,
$\boldsymbol{D}$ --- постоянный вектор,
$\boldsymbol{A}(\mathcal{P})$ --- отображение точки $\mathcal{P}$ в
якобиан кривой голоморфными интегралами \cite{5}.

В первой же работе 1974 г. \cite{19} Новиковым была выявлена
ключевая роль высших стационарных уравнений КдВ и оператора,
коммутирующего с оператором Шрёдингера. Помимо того, что в ней были
поняты и <<алгебраизированы>> глубокие результаты спектральной
теории, там было обнаружено, что уравнения, которые сейчас принято
называть уравнениями Новикова, являются конечномерными
гамильтоновыми вполне интегрируемыми динамическими системами, а
коэффициенты возникающей алгебраической кривой --- их интегралами
\cite{39}. Позже Кричевер показал \cite{25}, что наличие парного
коммутирующего оператора эквивалентно конструкции \reff{theta}
вместе со спецификацией алгебраической кривой. Это --- универсальное
свойство конечнозонных операторов вообще и оно может быть положено в
основу определения таковых \cite{2}. Как обнаружил  в 70-х годах
Кричевер \cite{24}, коммутирующие операторы и даже их связь с
абелевыми функциями исследовались еще в 20-х годах \cite{31,32}.

\subsection{Краткое содержание и мотивации}

Интенсивное исследование уравнения Шрёдингера (оператора Хилла) за
последние десятилетия привело к тому что теория его интегрирования
почти канонизирована. В этой связи не безинтересно привести новые
результаты
. Этому посвящен \S\,2. В
\S\,3 обсуждается соотношение между конечнозонными и более общими
интегрируемыми уравнениями. Уравнение Шрёдингера является примером
весьма специального вида, причем далеко не характерным для
обобщений. Последующие параграфы \S\,4--7 посвящены построению
необходимых атрибутов интегрируемости в этих случаях. К ним
относятся формулы следов, уравнения Дубровина и задача обращения
абелевых интегралов. Поскольку вопрос о переносе этих построений на
негиперэллиптические задачи до недавнего времени не рассматривался,
\S\,8--9 заполнены примерами.

Автор выражает признательность С.\,Цареву, В.\,Кузнецову и
В.\,Энольскому за  обсуждения и консультации, а проф. К.\,Эйлбеку и
Э.\,Превиато за интерес к работе.

\section{Интегрируемость уравнения Шрёдингера}
\noindent В 1994 г. \cite[стр.\,84--85]{38} Матвеев обратил внимание
на старую работу Драшa  \cite{49_2}, где в чрезвычайно сжатой форме
содержатся важнейшие результаты  по интегрированию уравнения
\begin{equation}\label{schr}
\Psi''-u(x)\,\Psi=\lambda\,\Psi\,.
\end{equation}
Драш построил  потенциалы и решения $\Psi$, когда \reff{schr}
интегрируются в неопределенных квадратурах и объявил что это все
возможные случаи.  Раннее упоминание его работ относится к началу
80-х годов, однако не указывается на  отличие его идеологии от
спектрального и $\Theta$-функционального подходов. Поскольку в
работах \cite{49_1,49_2} отсутствуют какие-либо пояснения и
доказательства, было бы уместно привести соответствующее изложение
теории и ряд моментов, которые обходятся стороной в  литературе.
Например, по каким причинам возникает фундаментальная функция $R$, а
появляющаяся алгебраическая кривая должна иметь конечный род, как
вводить переменные $\gamma_k^{}$ и почему возникают именно они?
Новая формула для $\Psi$-функции (формула \reff{final}), как
завершающий штрих к идеологии Драша, приводит к известному
$\Theta$-функциональному описанию теории.

\subsection{Интегрируемость Пикара--Вессио уравнения \reff{schr}}
Вопрос об  интегрируемости дифференциального уравнения в квадратурах
решается положительно, если группа его непрерывных симметрий
разрешима.  Используя эти факты (подробности см. в \cite{br4})
выводим формулу
\begin{equation}\label{pm}
\Psi_{\pm}(x;\,\lambda)=\sqrt{R(x;\lambda)}\,
\exp\!\!\int\limits^{\,\,x}\!\!\frac{\pm\mu\,dx}{R(x;\lambda)}\,,
\end{equation}
в которой функция $R=R(x;\lambda)$ удовлетворяет дифференциальному
уравнению
\begin{equation}\label{hermit}
R'''-4\,(u+\lambda)\,R'-2\,u'\,R=0\,.
\end{equation}
Интегрируя его один раз, и обозначая константу интегрирования как
$\mu$,
получаем
\begin{equation}\label{muR}
\mu^2=-\mfrac12\,R\,R''+\mfrac14\,R'^2+(u+\lambda)\,R^2\,.
\end{equation}
Вариации  формулы \reff{pm} многократно появлялись в современной
литературе \cite{8,29,39,38}, но точный смысл --- в \cite{49_2}.
Наличие \reff{pm}, как впрочем и дальнейшей \reff{exp},  не означает
никакой интегрируемости. Это анзац, так как уравнение \reff{schr}
имеет решение вида \reff{pm} при любых $u(x)$, не только в
конечнозонном классе. Все зависит от потенциала, а решение уравнения
\reff{hermit} следует предъявить явными интегралами, иначе процедура
сведется к переобозначениям.  Более того, интегрирование методами Ли
не всегда возможно\footnote{Отсутствие точечных симметрий не
означает неинтегрируемость. Контрпример
--- уравнение \reff{spectrsk}.}. С другой стороны, сам факт существования квадратурного
представления для интегралов линейных уравнений точно
устанавливается  через теорию Пикара--Вессио \cite{kolchin,kapl} и
ее современные разработки \cite{singerput}.

В дифференциальной теории Галуа уравнение \reff{hermit} называется
2-й симметрической степенью уравнения \reff{schr} \cite{singer}.
Желая ввести дифференциальное поле, над которым разворачивается
интегрирование, мы получаем, из формулы \reff{pm}, что этому полю
должна принадлежать функция $R(x;\lambda)$. Она, как известно,
является бесконечным рядом по $\lambda$ и дифференциальным полиномом
по $u(x)$. Это известные бесконечно зацепляющиеся рекуренции
Гельфанда--Дикого \cite{28, dik4}. Отсюда следует, что условием
интегрируемости в квадратурах уравнения \reff{schr} является
конечность этого ряда. Уравнение \reff{hermit} должно иметь хотя бы
одно решение, полиномиальное по $\lambda$. Это приводит к факту
появления (без явного описания) поля с дифференцированием
$\partial=\partial_x$ и условиям на потенциал. Эти условия
дифференциальны и известны как уравнения Новикова. Будем называть
это базовое дифференциальное поле, полем Новикова
$\mathcal{N}_\partial(u,\lambda)$  в $u(x)$-представлении, имея в
виду то, что $u(x)$ является решением обыкновенного
дифференциального уравнения Новикова порядка $2g+1$. В подполе
констант $\mathcal{N}_\partial$, помимо $\lambda$, также входят
константы, происходящие от рекурсии Гельфанда--Дикого и другие точки
ветвления кривой \reff{muR}. Сопоставляя с известными фактами,
заметим что абелевы функции образуют дифференциальное поле, только
если точка якобиана линейно зависит от переменной $x$. В дальнейшем,
во избежание постоянного упоминания такого типа сечения якобианов,
мы будем подразумевать так всегда, когда речь идет о многообразиях
Якоби или абелевых функциях на них. Функции $u(x)$, как решения
уравнений Новикова, являются такими абелевыми функциями.
Квадратурная интегрируемость уравнения \reff{schr} выражается в виде
следующего утверждения.

{\bf Теорема 1.} {\em Расширение Пикара--Вессио
$\mathcal{N}_\partial(u,\lambda)\langle\Psi_{\pm}\rangle$ является
Лиувиллевским расширением со степенью трансцендентности один.
Соответствующая дифференциальная группа Галуа
$\mathrm{Gal}\big(\mathcal{N}_\partial(u,\lambda)\langle\Psi_{\pm}\rangle\big)$
связна и сопряжена группе
$\mathfrak{G}=\Big(\begin{smallmatrix}\alpha & 0_{}\\0 &
\alpha^{-1}\end{smallmatrix}\Big)$, где $\alpha\in\mathbb{C}$}.

\noindent {\em Доказательство\/}. Лиувиллевость и трансцендентность
расширения очевидна, так как интеграл \reff{pm}, в случае общего
положения\footnote{Произвольное $\lambda$. Это фундаментальное
требование подразумевается повсюду и больше не оговаривается.}, не
принадлежит $\mathcal{N}_\partial(u,\lambda)$. Выберем канонический
базис решений в виде \reff{pm}. Из него получаем соотношения на
величины $\{\Psi_{\pm},\Psi_{\pm}'\}$:
$$
 \Psi_{-}=\frac{R}{\Psi_{+}}\,,\qquad
\Psi_{+}'=\frac{R'+2\mu}{2R}\,\Psi_{+}\,, \qquad
\Psi_{-}'=\frac{R'-2\mu}{2\Psi_{+}}\,,
$$
инвариантность которых относительно  линейного преобразования
$\mathfrak{G}=\big(\begin{smallmatrix}\alpha & \beta\\\gamma &
\delta\end{smallmatrix}\big)$ базиса $\Psi_{\pm}$ полностью
определяет группу
$\mathrm{Gal}\big(\mathcal{N}_\partial(u,\lambda)\langle\Psi_{\pm}\rangle\big)$.
Проверяя, получаем матрицу, фигурирующую в утверждении теоремы. Из
первого соотношения следует что  одно из решений алгебраически
определяется через другое: степень трансцендентности понижается до
единицы. \hfill$\blacksquare$

{\em Замечание\/}. В доказательстве нигде не использовался вид
дифференциального полинома $R$. Теорема справедлива для любых
интегрируемых  $\lambda$-пучков порядка 2. Например для задач,
связанных с нелинейным уравнением Шрёдингера или уравнением
sin-Гордон.

К этому моменту мы получили лишь необходимые условия полной
интегрируемости. Достаточность требует явного предъявления полей
$\mathcal{N}_\partial(u,\lambda)$. Это тоже проблема интегрирования,
но интегрирования нелинейных уравнений --- уравнений Новикова.  Хотя
они и могут быть представлены как интегрируемые по Лиувиллю
гамильтоновы системы \cite{39,dik4}, необходимости их интегрирования
таким способом (см. \cite[\S\,6--7]{27}, \cite{58, dik4}), в
рассматриваемом нами контексте теории Пикара--Вессио, можно
избежать. Половина констант интегрирования
--- это кривая \reff{muR}. Оставшееся следует из
формул следов и задачи Якоби.

\subsection{Процедура интегрирования. Спектральный и
квадратурный подходы}

В \cite{br4} мы прокомментировали то, что, по всей видимости, имел в
виду Драш, когда вводил новые переменные $\gamma_k^{}$ как корни
фундаментального полинома
\begin{equation}\label{factor}
R(x;\lambda)=(\lambda-\gamma_1^{}(x))\cdots
(\lambda-\gamma_g^{}(x))\,.
\end{equation}

{\bf Теорема 2.} {\em  Квадратурный  и спектральный подход
эквивалентны. Квадратурная версия спектральных $\Psi$-формул
Ахиезера--Матвеева--Итса \cite{15,21} имеет вид
\begin{equation}\label{final}
\begin{array}{l}
\displaystyle \Psi_{\pm}(x;\lambda)=\mbox{\rm\large exp}\,
\frac{1}{2}\!\left\{\! \int\limits^{\,\gamma_{{}_{1}}\!(x)}
\!\frac{w\pm\mu}{z-\lambda}\,\frac{dz}{w}\,\,\,\,+ \cdots +
\int\limits^{\,\gamma_{{}_{g}}\!(x)}
\!\frac{w\pm\mu}{z-\lambda}\,\frac{dz}{w} \right\}\;.
\end{array}
\end{equation}
Явный переход осуществляет теорема Вейерштрасса о перестановке
аргументов и параметров в  нормализованных абелевых интегралах 3-го
рода на гиперэллиптических кривых\/}.

\noindent {\em Доказательство\/}. Формула \reff{final} получается
подстановкой \reff{factor} в \reff{pm} с последующей заменой
интегрирования по $x$ на интегрирование на кривой $w^2=(z-E_1)\cdots
(z-E_{2g+1})$ абелевых дифференциалов 3-го рода в силу
дифференциальных уравнений Дубровина \cite{49_2,7}
\begin{equation}\label{dubrovin}
\frac{d\gk}{dx}=2\frac{\sqrt{(\gk-E_1)\cdots
(\gk-E_{2g+1})}}{\displaystyle\prod\limits_{j\ne k}(\gk-
\gamma_j^{})}\,.
\end{equation}
Далее, подстановка формулы \reff{final}  в уравнение \reff{schr}
дает тождество, если воспользоваться формулой следов
$u=2\sum\gk-\sum E_k$ \cite{49_2, 21} и уравнениями \reff{dubrovin}.
Комбинирование правых частей в уравнениях \reff{dubrovin} позволяет
переписать их в виде неопределенных квадратур \cite{49_2,7}
\begin{equation}\label{jac}
\sum\limits_{k=1}^{g}\int\limits^{\;\gamma_k^{}}\!z^{g-1}\,\frac{d
z}{w} = d_g+2\,x\,,\qquad
\sum\limits_{k=1}^{g}\int\limits^{\;\gamma_k^{}}\!z^n\,\frac{d z}{w}
= d_{n+1}\,,\quad n=0,1,\ldots,g-2\,.
\end{equation}
Оставшуюся часть доказательства, о перестановке аргументов и
параметров, см. в \cite[\S\,7]{br3}. \hfill$\blacksquare$

Мы приходим, таким образом, к естественному выводу что
рассматриваемые концепции интегрируемости (линейная и нелинейная)
редуцируются к интегрируемости Пикара--Вессио в ее простейшей форме:
{\em <<конечнозонные>>  группы
$\mathrm{Gal}\big(\mathcal{N}_\partial(u,\lambda)\langle\Psi_{\pm}\rangle\big)$
всех этих уравнений диагональны, не зависят от параметра в
уравнении, необходима {\em единственная\/} квадратура, а операторы
\reff{schr} факторизуемы над полями
$\mathcal{N}_\partial(u,\lambda)$\/}.

Отметим принципиальное отличие между спектральным и квадратурным
способами получения формул. К этому моменту не возникал не только
анализ на римановых поверхностях, но и они сами. Имеется лишь
обозначение $\mu=\sqrt{(\lambda-\gamma_1^{})\cdots
(\lambda-E_{2g+1})}$\,. В тоже время в спектральном подходе
Ахиезера--Матвеева \cite{15,4} существенно используется введение и
нормализация абелевых интегралов всех родов. Без этого, в частности,
важная формула Матвеева--Итса (4.12--13) в \cite{21}, которая
заслуживает большего внимания чем есть, быть получена не может.

\subsection{$\Theta$-функции}
Следующим пунктом является проблема аналитического представления для
предыдущих формул (Матвеев--Итс (1975) \cite{21}), которые отражают
свойство задачи быть интегрируемой не зависимо от выбора переменных
или методов вывода.  Поскольку расширение трансцендентно, это могут
делать только новые функции, которые с неизбежностью должны быть
введены (см. \S\,9 в \cite{br4}). Например, как многомерные
$\Theta$-ряды \cite{21,38}. Такой естественный порядок действий,
применительно к спектральным задачам, определяемыми обыкновенными
диференциальными уравнениями, и (1+1)-интегрируемым моделям,
позволяет не привлекать $\Theta$-функции в качестве отправной точки
при интегрировании.

{\bf Теорема 3.} {\em $\Theta$-функциональное представление
\reff{theta} для $\Psi(x;\lambda)$ и  \reff{its} для потенциала
$u(x)$ выводимо из квадратурной формулы \reff{final}. Выражения
\reff{theta} и \reff{final} пропорциональны\/}.

\noindent {\em Доказательство\/} и, в частности, появление
мероморфного  интеграла $\Omega(\mathcal{P})$  и его  периодов
$\boldsymbol{U}$, см. в \cite[\S\,7]{br3}. Иными словами (и не
приводя необходимых оговорок), поля Новикова
$\mathcal{N}_\partial(u,\lambda)$  являются и абелевыми и
интегрируемыми. Лиувиллевские расширения
$\mathcal{N}_\partial(u,\lambda)\langle\Psi_{\pm}\rangle$ выражаются
в $\Theta$-функциях как и сами элементы полей. В этом смысле, через
$\Theta$-функции реализуется шаг, который следует рассматривать как
следующий после получения собственно значков интеграла в \reff{pm} и
\reff{jac}: интегралы и дифференциалы от  $\Theta$-отношений
выражаются снова в $\Theta$. В вырожденном солитонном/рациональном
случае, когда кривая \reff{muR} имеет нулевой род, такая замкнутость
реализуется через линейную функцию $x$ и экспоненту от нее. Во всех
остальных интегрируемых случаях, замыкание осуществляют
$\Theta$-функции с {\em линейной\/} зависимостью от $x$ и экспоненты
Кричевера (см. формулу \reff{theta}).

\subsection{$\Theta$-представление полей Новикова}
Теорема 3 говорит о том, что <<$\Theta$-функциональное
интегрирование>> является  <<$\Theta$-функциональным
представлением>> для инвариантного свойства уравнения быть
интегрируемым по Пикару--Вессио и подсказывает искать представления
дифференциальных полей Новикова $\Theta$-функциями, а потом
расширять их. Тоже в $\Theta$. Для эллиптических потенциалов это
действительно возможно.

Пусть $\thetaAB{\alpha}{\beta}$ --- стандартные $\theta$-ряды Якоби
с характеристиками:
$$
\thetaAB{\alpha}{\beta}(x|\tau)= \sum\limits_{{}^{\scriptscriptstyle
k=-\infty}}^{\scriptscriptstyle \infty} e^{\pi i
\left(\!k+\frac\alpha2\!\right)^{\!2}\tau+ 2\pi
i\left(\!k+\frac\alpha2\!\right)\!\left(\!x+\frac\beta2\!\right)}\,.
$$
Пусть $\vartheta\equiv\theta(0|\tau)$ --- $\vartheta$-константы и
$\theta_{1\!\!\!}'$ обозначает производную по $x$ от ряда
$-\thetaAB{1}{1}(x|\tau)$. Обозначим  период нормализованного
мероморфного эллиптического интеграла как $\eta=\zeta(1|1,\tau)$.

{\bf Теорема 4.} {\em Функции $\thetaAB{\alpha}{\beta}$  и
$\theta_{1\!\!\!}'$ с произвольными целыми характеристиками
$(\alpha,\beta)$ $($т.\,е. функции Якоби $\theta_{1,2,3,4}^{}$$)$
замкнуты относительно дифференцирования над
$(\vartheta,\eta)$-константами и удовлетворяют автономным
обыкновенным дифференциальным уравнениям\/}
\begin{equation}\label{new}
\left\{
\begin{array}{rl}
\ds \frac{\partial\thetaAB{\alpha}{\beta}}{\partial x}\!\!&=\ds\;
\frac{\theta_{1\!\!\!}'}{\thetaAB{1}{1}} \,\thetaAB{\alpha}{\beta}-
(-1)^{\alpha\left[\frac{\beta}{2}\right]}_{\mathstrut}\,
\pi\,\varthetaAB{\alpha}{\beta}^2\!\cdot\!
\frac{\thetaAB{1\mbox{\tiny--}\alpha}{0}\,
\thetaAB{0}{1\mbox{\tiny--}\beta}}{\thetaAB{1}{1}}\\\\
\ds \frac{\partial\theta_{1\!\!\!}'}{\partial x}\!\!&=\ds\;
\frac{\theta_{1\!\!\!}'^2}{\thetaAB{1}{1}}-\pi^2
\varthetaAB{0}{0}^2\,\varthetaAB{0}{1}^2 \!\cdot\!
\frac{\thetaAB{1}{0}^2}{\thetaAB{1}{1}}-
\Big\{4\eta+\mfrac{\pi^2}{3}\big(\varthetaAB{0}{0}^4+
\varthetaAB{0}{1}^4\big) \Big\} \!\cdot\!\thetaAB{1}{1}
\end{array}\right..
\end{equation}
{\em Дифференциальное расширение полей Новикова для эллиптических
потенциалов  до произвольных $\theta$-отношений  осуществляют
элементы $\theta_{1\!\!\!}'$\/:
$\mathcal{N}_\partial(u,\lambda)\subset
\mathbb{C}(\lambda,\mu,\vartheta,\eta;\theta,\theta_{1\!\!\!}')$}.

\noindent{\em Доказательство\/}. Это не что иное как
$\theta$-представление соотношений между функциями Вейерштрасса
$\zeta$ и $\wp$.  Произвольные характеристики выбраны для унификации
формул. Полиномиальные тождества между $\theta$-функциями,
дополненные уравнениями \reff{new}, соответствуют Вейерштрассовскому
базису $\sigma,\zeta,\wp,\wp'$. \hfill$\blacksquare$

Теорема 4 описывает  дифференциальные свойства  функций более
широкого класса чем абелевых  или функций типа БА. Это справедливо
не только для рода $g=1$, но и в случаях, когда $g$-мерный якобиан
допускает редукцию $\Theta$-функции на якобиевские.

Далее будем использовать обозначения  формулы \reff{theta}. Назовем
$\Theta(x\boldsymbol{U}+\boldsymbol{C})\,e^{kx}$, где $k$ и
$\boldsymbol{C}$ постоянны относительно $\partial$, {\em
линейно-экспоненциальным дивизором\/} ({\em линейным\/}, если $k=0$)
или $\boldsymbol{\mathcal{L}}_x$-дивизором. Образуем из таких
дивизоров поле над $\mathbb{C}$. Оно дифференциально, так как по
формуле Матвеева--Итса
\begin{equation}\label{its}
u(x)=-2\,\frac{d^2}{dx^2}\ln\Theta(x\,\boldsymbol{U}+\boldsymbol{D})
\,e^{kx}-\sum\limits_{j=1}^{2g+1}E_j\,,
\end{equation}
а сам потенциал удовлетворяет уравнениям Новикова. Их порядок
конечен.

Рассмотрим поле $\Theta_\partial$, порожденное одним линейным
дивизором: $\Theta_\partial\equiv
\mathbb{C}_\partial\big(\Theta(x\boldsymbol{U}+\boldsymbol{D})\big)$.
Имеем $ \mathcal{N}_\partial(u)\subset\Theta_\partial$. Хотя
элементами $\Theta_\partial$ являются уже не только абелевы функции,
это расширение вполне определено, так как $\Theta$-ряд и
$\boldsymbol{b}$-периоды $\boldsymbol{U}$ однозначно строятся по
$u(x)$.  По этой причине будем  считать что
$\mathcal{N}_\partial(u)=\Theta_\partial$, т.\,е. уравнение
\reff{schr} задано над $\Theta_\partial$, которое естественно
называть $\Theta$-{\em представлением\/} полей Новикова. Константу
$\lambda$  пока игнорируем.

\subsection{Интегрирование как
линейно-экспоненциальное $\Theta$-расширение}

Рассмотрим поля Пикара--Вессио
$\mathcal{N}_\partial(u,\lambda)\langle\Psi_{\pm}\rangle$ в
представлении $u(x)$. Их трансцендентность, как расширений
$\mathcal{N}_\partial(u,\lambda)$, трояка:  1) интеграл как
неалгебраическая операция в $\mathcal{N}_\partial(u,\lambda)$, 2)
существенно неалгебраическая зависимость $\Psi$ от параметра
$\lambda$  и 3) собственно присоединение трансцендентного элемента
(экспонента). В тоже время, на основании Теоремы 1 и  3, мы видим,
что поля имеют следующую структуру:
$$
\mathcal{N}_\partial(u,\lambda)\langle\Psi_{\pm}\rangle=
\mathbb{C}_\partial\Big(
\mfrac{\Theta\big(x\boldsymbol{U}+\boldsymbol{D}+\boldsymbol{A}(\mathcal{P})\big)}{
\Theta(x\boldsymbol{U}+\boldsymbol{D})}\,
e^{\Omega(\mathcal{P})x}\Big)\,.
$$
Отсюда следует, что в $\Theta$-представлении  интегрирование как
таковое исчезает, а  все пункты 1--3) сводятся к одной операции. К
полю, над которым задано уравнение, достаточно присоединить
трансцендентный элемент такого же вида, как и элемент, порождающий
само поле:
$$
\mathcal{N}_\partial(u)=\mathbb{C}_\partial\big(
\Theta(x\boldsymbol{U}+\boldsymbol{D})\big) \subset
\mathbb{C}_\partial\big( \Theta(x\boldsymbol{U}+\boldsymbol{D})\,
e^{kx},\,
\Theta(x\boldsymbol{U}+\boldsymbol{D}+\boldsymbol{A}(\mathcal{P}))\,
e^{\Omega(\mathcal{P})x}\big)=\mathcal{N}_\partial(u,\lambda)\langle\Psi_{\pm}\rangle\,.
$$
Важно отметить, что вместе с <<исчезновением линейного
интегрирования>> отпадает также первичная проблема построения
базового поля $\mathcal{N}_\partial(u)$ как проблема интегрирования
нелинейных уравнений Новикова. В $u(x)$-представлении она
определялась бы двумя трансцендентными операциями: интегралами от
алгебраической функции \reff{jac} и процедурой обращения. В
$\Theta$-представлении потенциал находится с помощью операций в поле
$\Theta_\partial$ по формуле \reff{its}.

{\bf Теорема 5.} {\em Пусть $u(x)$
--- конечнозонный потенциал. В $\Theta$-представлении,
интегрирование уравнения Шрёдингера \reff{schr} эквивалентно
умножению элемента $\Xi$, порождающего дифференциальное поле
$\mathcal{N}_\partial(u)$, на присоединяемый
линейно-экспоненциальный дивизор\/:}
\begin{equation}\label{lin}
\Psi_{\pm}\big(x;\lambda(\mathcal{P})\big)=
C_{\pm}\cdot\Xi(x)\cdot
\Theta\big(x\boldsymbol{U}+\boldsymbol{D} \pm
\boldsymbol{A}(\mathcal{P})\big) \, e^{\pm\Omega(\mathcal{P})x}\,.
\end{equation}

\noindent {\em Доказательство\/}. Возьмем
$\Xi(x)=\Theta(x\boldsymbol{U}+\boldsymbol{D})^{-1}$. В силу
\reff{its}  $\mathcal{N}_\partial(u)=\mathbb{C}_\partial(\Xi)$. Все
мероморфные интегралы на гиперэллиптических кривых меняют знак при
перестановке листов. \hfill $\blacksquare$

Теорема 5 говорит о том, что при переходе к $\Theta$-представлению,
уравнение \reff{schr} интегрируется, как
если бы оно было уравнением с {\em постоянными\/} коэффициентами%
\footnote{Это можно рассматривать как продолжение  известной
интерпретации интегрирования нелинейных уравнений {\sc мозр}, как
линеаризации [$L,A$]-парами. В свою очередь, в классе
солитонных/конечнозонных потенциалов, процедура интегрирования
тривиализируется при надлежащем  выборе представления для областей
интегрирования (областей рациональности).}. В таком простейшем
(0-зонном) случае мы имели бы аналог схемы \reff{lin}:
$\Psi_{\pm}(x;\lambda)= C_{\pm}\cdot 1\cdot e^{\pm a(\lambda)x}$.
Заметим, что структура решения \reff{lin} в виде простого
перемножения элементов порождающих $\mathcal{N}_\partial(u)$ и его
расширение $\mathcal{N}_\partial(u)\langle\Psi_{\pm} \rangle$  не
является общей для уравнений интегрируемых присоединением линейных
экспонент или, тем более, для уравнений с просто разрешимой группой
Галуа. Появление такого свойства обязано наличию в уравнении
$\lambda$.

Если расширение алгебраично по $\lambda$, то присоединение
трансцендентного элемента  имеет место в любом представлении,  так
как степень трансцендентности равна единице (теорема 1). С другой
стороны, в соответствии с теоремой 3,
$\boldsymbol{\mathcal{L}}_x$-дивизор есть фактически результат
квадратур. Переход от $u$- к $\Theta$-представлению трансцендентен
по константам поля ($B$-матрицы кривых, $\boldsymbol{U}$-периоды и
вектор $\boldsymbol{D}$). По этой причине мы можем  рассматривать
уравнение \reff{schr} заданным над $\lambda$-пучком (полем)
$\boldsymbol{\mathcal{L}}_x$-дивизоров
$\Theta_\partial\big(\lambda(\mathcal{P})\big)\equiv
\mathbb{C}_\partial\big(\Theta(x\boldsymbol{U}+\boldsymbol{D}+
\boldsymbol{A}(\mathcal{P})) \, e^{\Omega(\mathcal{P})x}\big)$,
т.\,е. считать $\mathcal{N}_\partial(u,\lambda)=
\Theta_\partial\big(\lambda(\mathcal{P})\big)$. Хотя такое поле
порождено бесконечным количеством элементов, само уравнение
\reff{schr} тоже есть бесконечный $\lambda$-пучок уравнений. Любой
$\boldsymbol{\mathcal{L}}_x(\mathcal{P})$-дивизор порождает
некоторое решение уравнения Новикова, т.\,е. потенциал. Его можно
фиксировать величиной $\boldsymbol{D}+\boldsymbol{A}(\mathcal{P})$:
$\mathcal{N}_\partial(u)=\mathbb{C}_\partial\big(
\Theta(x\boldsymbol{U}+\boldsymbol{D}_0+
\boldsymbol{A}(\mathcal{P}_0)) \, e^{\Omega(\mathcal{P}_0)x}\big)$.
Параметр $\lambda$, при такой формулировке, как константу поля,
следует рассматривать как квадрат частного
$\boldsymbol{\mathcal{L}}_x$-дивизоров взятых в точке $x=0$. В самом
деле, $\lambda$, как мероморфная функция на кривой \reff{muR}, имеет
представление в виде $\lambda(\mathcal{P})\sim \Theta^2
\raisebox{0.04em}{\mbox{$ \scalebox{0.5}[0.9]{\big[}
\!\!\raisebox{0.1em} {\scalebox{0.8}{\mbox{\tiny
$\begin{array}{c}\boldsymbol{\alpha}\\\boldsymbol{\beta}\end{array}$}}}
\!\! \scalebox{0.5}[0.9]{\big]}$}}(\boldsymbol{A}(\mathcal{P}))
\big/\Theta^2 \raisebox{0.04em}{\mbox{$ \scalebox{0.5}[0.9]{\big[}
\!\!\raisebox{0.1em} {\scalebox{0.8}{\mbox{\tiny
$\begin{array}{c}\boldsymbol{\gamma}\\\boldsymbol{\delta}\end{array}$}}}
\!\! \scalebox{0.5}[0.9]{\big]}$}}(\boldsymbol{A}(\mathcal{P}))$.
Это было показано Примом во 2-м издании его  диссертации
\cite{prym}. Окончательно, мы приходим к тому, что в
$\Theta_\partial\big(\lambda(\mathcal{P})\big)$-представлении
исчезает и присоединение трансцендентного элемента, а группа Галуа
становится тривиальной: интеграл уравнения \reff{schr} есть
отношение двух элементов поля.

Для  замыкания картины нам следовало бы ответить на вопрос о природе
происхождения $\Theta$-ряда. Здесь мы ограничимся  лишь тем
замечанием что, хотя постулируемая форма $\Theta$-ряда не может
объяснять какую-либо интегрируемость, сам ряд (его общий член) может
быть получен регулярной процедурой. Этот факт (и его следствия)
является важным, но его описание выходит за рамки настоящей работы и
будет предметом отдельного изложения.

\subsection{Дифференциальная замкнутость линейных
$\boldsymbol{\mathcal{L}}_x$-дивизоров}

Пусть $g$-мерная функция $\Theta(x\boldsymbol{U}+\boldsymbol{D})$
редуцируется к функциям Якоби $\theta_{1,2,3,4}$ с надлежащими
фазами. Теорема 4 означает тогда, что в таких ситуациях механизм
интегрирования содержится в дифференциальной замкнутости функций
Якоби относительно их первого аргумента. Учитывая то что
изоспектральные деформации интегрируемых (1+1)-уравнений тоже
описываются линейными фазами
$x\boldsymbol{U}+t\boldsymbol{V}+\boldsymbol{D}$ на якобианах,
получаем, что все такие точно решаемые линейные/нелинейные модели,
включая уравнения в частных производных, являются следствиями
фундаментальной системы обыкновенных дифференциальных уравнений
\reff{new} и полиномиальных тождеств между $\theta$-функциями. Этот
факт говорит о том, что между лиувиллевостью конечнозонных
расширений Пикара--Вессио, т.\,е. присоединением  экспонент от {\em
однократных\/} интегралов, и дифференциальной замкнутостью
$\theta,\theta'$-функций {\em первого порядка\/}, имеется своего
рода изоморфизм. Одно влечет другое. Будет ли подобное в случае
общего положения? Дифференциальные свойства общих $g$-мерных
$\Theta$-функций выражаются, через абелевы $\Theta$-отношения,
соотношениями между их {\em частными\/} производными в виде
интегрируемых уравнений \cite{baker, BEL}, в то время как их
эллиптические редукции --- обыкновенным дифференциальным уравнением
$\wp'^2=4\wp^3-g_2^{}\wp-g_3^{}$. Определяются ли тогда
дифференциальные свойства сечений общих якобианов типа
$x\boldsymbol{U}+\boldsymbol{D}$ конечным набором {\em
обыкновенных\/} дифференциальных уравнений порядка один
--- уравнениями на функции $\Theta$ и $\Theta'$?

Более точная формулировка могла бы выглядеть следующим образом.
Пусть $\boldsymbol{U}$  --- вектор $\boldsymbol{b}$-периодов
нормализованного абелева интеграла 2-го рода на гиперэллиптической
кривой, имеющего единственный полюс 1-го порядка. Пусть
$\big\{\Theta \raisebox{0.04em}{\mbox{$ \scalebox{0.5}[0.9]{\big[}
\!\!\raisebox{0.1em} {\scalebox{0.8}{\mbox{\tiny
$\begin{array}{c}\boldsymbol{\alpha}\\\boldsymbol{\beta}\end{array}$}}}
\!\!
\scalebox{0.5}[0.9]{\big]}$}}(x\boldsymbol{U}+\boldsymbol{D}),\,
\Theta' \raisebox{0.04em}{\mbox{$ \scalebox{0.5}[0.9]{\big[}
\!\!\raisebox{0.1em} {\scalebox{0.8}{\mbox{\tiny
$\begin{array}{c}\boldsymbol{\alpha}\\\boldsymbol{\beta}\end{array}$}}}
\!\!
\scalebox{0.5}[0.9]{\big]}$}}(x\boldsymbol{U}+\boldsymbol{D})\big\}$
обозначает $\Theta$-функции с целочисленными характеристиками.
Являются ли они дифференциально замкнутыми над полем
$(\boldsymbol{\vartheta},\boldsymbol{U})$-констант, т.\,е.
существует ли аналог уравнений \reff{new}?  Другими словами,
является ли это свойство, скорее чем уравнение теплопроводности (как
уравнение в частных производных), характеристическим  для
$\Theta$-функций и прямолинейных сечений якобианов, которые
возникают в общей теории солитонов?

Обратная, хотя и более широкая, постановка этих вопросов,
представляется не менее интересной. Допустимо ли вообще
интегрировать  абелевы функции в том смысле, что неопределенные
интегралы от произвольных абелевых функций на многообразиях Якоби
алгебраических кривых, выражались бы  через частные линейных
дивизоров, логарифмы от них и функции $\Theta'$. Примеры, когда это
так, многочисленны. Спрашивается, всегда ли это так и, если нет, то
что необходимо для интегрального замыкания?

\subsection{Алгоритмы и эллиптические солитоны}
Некоторые примеры эллиптических конечнозонных потенциалов
\cite{smirnov} уже встречались в рамках алгоритмического
интегрирования линейных уравнений \cite{singer}. Однако, насколько
нам известно, ключевые объекты алгебро-ге\-о\-мет\-ри\-чес\-ко\-го
интегрирования (переменные $\gk$ и $\Theta$-функции),  не
упоминаются в соответствующей литературе (см. например  монографии
\cite{singerput}, \cite{berkovich}, \cite{ispanec}). Покажем как
генерировать широкий класс  решаемых уравнений с рациональными и
алгебраическими коэффициентами.

Пусть $u(x)$ --- эллиптический конечнозонный потенциал, а
$u=Q_1^{}(\wp)+Q_2^{}(\wp)\,\wp'$ есть его Вейерштрассовское
представление. Сделаем замену переменных $x \to \wp,\;\Psi \to
\widetilde\Psi= \sqrt{\mathstrut \wp'}\,\Psi. $ Уравнение
\reff{schr} примет вид
\begin{equation}\label{kov}
\widetilde\Psi_{\mathit{\wp\wp}}+
\left\{3\,\frac{(\wp^2+\frac14g_2^{})^2+2\,g_3^{}\wp
}{(4\wp^3-g_2^{}\wp-g_3^{})^2}
-\frac{Q_1^{}(\wp)+Q_2^{}(\wp)\,\wp'+\lambda}{4\wp^3-g_2^{}\wp-g_3^{}}
 \right\}\widetilde\Psi=0\,.
\end{equation}
По предыдущим построениям все уравнения \reff{kov} интегрируемы в
квадратурах и факторизуемы в эллиптических функциях. Достаточно
сделать упомянутую замену переменных в формуле \reff{pm}. Она
превратится в эллиптический интеграл.  В то же время известные
алгоритмы  Зингера и Ковачика \cite{singer,kov} доставляют
самостоятельные процедуры интегрирования таких уравнений. Это дает
пользу для обоих методов. Уравнения с рациональными коэффициентами
получаются если рассматривать четные потенциалы, т.\,е. с полюсами в
нуле (уравнения Ламе) и полупериодах (Дарбу, Трейбиш--Вердье и др.).
Поскольку конечнозонные потенциалы исчерпывают все интегрируемые
случаи мы имеем следующее.

{\bf Предложение 1.} {\em Алгоритмы Зингера--Ковачика, будучи
примененные к интегрируемым уравнениям вида
$$
\frac{\Psi''}{\Psi}-\lambda=\mbox{\rm функция  от \ $x$}\,,
$$ вкладываются в теорию конечнозонного интегрирования
как частный случай\/}.

Таким образом, формулы Итса--Матвеева для $\Psi$ вида \reff{theta} и
потенциалов вида \reff{its} доставляют общий ответ и в контексте
данных алгоритмических методов. Неэллиптические  потенциалы следует
рассматривать как расширения алгебраических \reff{kov} либо,
отталкиваясь от предыдущего пункта, естественно ожидать, что
$\Theta$-функции с целыми характеристиками реализуют
<<алгебраизацию>> интегрирования. Эллиптический случай соответствует
таким редукциям Якобиевых многообразий (они являются
алгебраическими), что в качестве образующих полей, вместо всего
множества необходимых $\big\{\Theta \raisebox{0.04em}{\mbox{$
\scalebox{0.5}[0.9]{\big[} \!\!\raisebox{0.1em}
{\scalebox{0.8}{\mbox{\tiny
$\begin{array}{c}\boldsymbol{\alpha}\\\boldsymbol{\beta}\end{array}$}}}
\!\! \scalebox{0.5}[0.9]{\big]}$}},\, \Theta'
\raisebox{0.04em}{\mbox{$ \scalebox{0.5}[0.9]{\big[}
\!\!\raisebox{0.1em} {\scalebox{0.8}{\mbox{\tiny
$\begin{array}{c}\boldsymbol{\alpha}\\\boldsymbol{\beta}\end{array}$}}}
\!\! \scalebox{0.5}[0.9]{\big]}$}}\big\}$, можно использовать только
две (любые образующие поля эллиптических функций) или даже одну
(уравнения с рациональными коэффициентами).

Точная интегрируемость представляет собой достаточно жесткое, но
конструктивное свойство. В него не попадают  все известные примеры
потенциалов в элементарных функциях с решениями в классе специальных
функций и их вырождений в элементарные (гармонический осциллятор,
ортогональные полиномы и т.\,д.). Они соответствуют кривым
бесконечного рода.

В заключение параграфа сделаем замечание связанное с кажущимся
противоречием. Квадратуры, т.\,е. появление $\Psi'/\Psi$, есть
локальное (по $x$) свойство уравнения, в то время как спектральная
конечнозонность  является существенно глобальным свойством
потенциала. Разрешение состоит в том, что локальная интегрируемость
требуется  при всех значениях $\lambda$, что ведет к появлению
уравнений Новикова. Аналитичность потенциала по $x$ существенна и
всюду подразумевается. Например кусочно-постоянный потенциал не
является конечнозонным.

\section{Интегрируемые и конечнозонные линейные уравнения}

\subsection{Интегрируемые уравнения}
В случае произвольных спектральных задач, группы точечных симметрий
Ли, вообще говоря, не приводят к успеху поскольку, даже если они
есть, они не обязаны быть разрешимыми. Однако, следуя предыдущему
параграфу естественно потребовать разрешимость соответствующей
группы Галуа, как алгебраической группы \cite{kolchin}. По крайней
мере это соответствовало бы формализованному понятию
<<интегрируемости в квадратурах>>, что и является предметом
рассмотрения. Естественно положить, что такая постановка должна
согласовываться с уже общеизвестными фактами. В частности: 1)
конечнозонные потенциалы генерируются стационарными иерархиями
[$L,A$]-пар нелинейных интегрируемых уравнений; 2) они являются
абелевыми функциями на якобианах алгебраических кривых; 3)
$\Theta$-аксиоматика функции Бейкера--Ахиезера, включая
(2+1)-уравнения, которые генерируют интегрируемые (1+1)-уравнения. С
другой стороны, эти признаки могут быть слишком ограничительными,
поскольку в такой формулировке мы отбрасываем ситуации, для которых
нет оснований считать уравнение <<неинтегрируемым>>. В самом деле,
интегрируемость, как инвариантное свойство, определяется независимо
от координат. Замены переменных, если они не выходят за <<пределы
квадратур>>, тоже порождают интегрируемые модели. Ближайшие примеры
доставляют фуксовы уравнения \reff{kov}, а из серии (2+1)-уравнений
показательным примером является цилиндрическое уравнение КП
(уравнение Джонсона) \cite{lip}. Следующий пример не является
искусственным и возникает в приложениях \cite{79}:
$$
\Psi''=\frac{\lambda}{u^2}\,\Psi\,.
$$
Конечнозонная теория для него в рамках \S\,2 остается без изменений.
Приведем ответы в случае $g=1$. $\Psi$-функция имеет вид \reff{pm},
в которой
$$
R(x;\,\lambda)=u\,\lambda+\mfrac14\,u^2\,u''-
\mfrac18\,u\,u'^2-\mfrac12\,\alpha\,u\,,
$$
а алгебраическая кривая \reff{muR} выглядит следующим образом:
$\mu^2=\lambda^3-\alpha\,\lambda^2-E_1\,\lambda-E_2$, где
$$
\left\{
\begin{array}{l}
E_1=\mfrac18\,u^3\,u^{\sss (\mathrm{IV})} +\mfrac14\,u^2\,u'\,u'''+
\mfrac {1}{64_{{}_{\mathstrut}}}\, (2\,u\,u'' -u'^2 -4\,\alpha)\,
(6\,u\,u''-3\,u'^2 + 4\,\alpha)\\
E_2=\mfrac {-1^{{}^{\mathstrut}}}{64}\,u^4\,u'''^2-\mfrac {1}{2^9}\,
\big(2\,u\,u''-u'^2 -4\,\alpha\big)\big((2\,u\,u''-u'^2)^2 -
16\,\alpha^2-64\, E_1{}^{{}^{\mathstrut}}\big)
\end{array}
\right.\,.
$$
В этом примере задача сводится к классической \reff{schr} с помощью
подстановки Лиувилля,  но  эволюция (по $x$) на якобиане кривой
будет нелинейна \cite{dmitr} и $\Psi$ не есть функция типа
\reff{theta}. Другими примерами $\Theta$-функционально интегрируемых
моделей не укладывающихся  в классическую аксиоматику функций БА,
являются деформации уравнений главного кирального поля \cite{bur} и,
в частности, уравнение Эрнста \cite{korotkin}. В этих задачах
коэффициенты $[U,V]$-пар, как функции от спектрального параметра,
имеют полюса зависящие от $x$ и сама кривая тоже зависит от $x$.
Детали  и дополнительные ссылки см., например, в работе
\cite{korotkin}. Далее, поскольку определение области интегрирования
$\mathcal{N}_\partial$ не является предметом $\Theta$-функциональных
методов, допустимо порождать интегрируемые задачи любым способом,
отличным от классической конечнозонной структуры, которая
описывается теоремой 5 и формулой \reff{theta}. В этом случае мы
можем, не меняя дифференциального поля, а точнее его расширения,
поменять структуру решения \reff{theta}. Например взяв такое
уравнение:
$$
\Psi''=\big\{6\,\wp(x)+2\,\wp(x-\lambda)+4\,\wp(\lambda)\big\}\Psi\,,
\qquad \Psi(x;\lambda)= \frac{\sigma^2(x-\lambda)}{\sigma^2(x)}\,
e^{2\zeta(\lambda)x}_{}\,.
$$
Примеры можно увеличивать даже не прибегая к
$\Theta$-представлениям. Ограничимся тем, что сама резольвента
задачи \reff{schr}, т.\,е. оператор \reff{hermit}, является таким
примером, интегрируемым в квадратурах с базисом решений
$\big\{\Psi_+^2,\,\Psi_+\Psi_-, \,\Psi_-^2\big\}$.  В этом случае,
как не трудно вычислить, группа Галуа не только диагональна, но и
дополнительно упрощается в группу сопряженную с группой
$\mathfrak{G}=\mathrm{Diag}(1,\alpha,\alpha^{-1})$, поскольку
существует решение (это $R=\Psi_+\Psi_-$) в поле
$\mathcal{N}_\partial$.

При таких или  подобных обобщениях нелинейные уравнения Новикова,
задачи обращения и другие  атрибуты точно интегрируемых моделей
становятся либо опосредованными или вовсе исчезают, но сам список
решаемых уравнений значительно расширяется. Важно заметить однако,
что если обобщение или даже прямая формулировка привлекают
$\Theta$-ряды необходимо отслеживать тот факт, что эти ряды должны
появляться из формул с неопределенными квадратурами в духе \S\,2.
Иными словами, $\Theta$-ряды, как $\Theta$-представления полей, не
должны возникать как специальные функции, обозначающие решение
задачи, наподобие того как  происходят интегрирования уравнений типа
Бесселя, Эйри или гипергеометрического уравнения. Мы не будем дальше
затрагивать возможную аксиоматику на основе квадратур с последующей
классификацией, а ограничимся теми уравнениями, где линейная и
нелинейная интегрируемость возникают и замыкаются друг на друге. Это
освобождает нас от привлечения $\Theta$-методов в последующих
параграфах. В них мы рассмотрим иерархии скалярных спектральных
задач, которые известны как иерархии Гельфанда--Дикого
\cite{58,59,dik4}.

С точки зрения квадратурной интегрируемости, формул следов и задачи
Якоби такие уравнения до недавнего времени не рассматривались, хотя
другие объекты их конечнозонного интегрирования известны. Например
для уравнения Буссинеска аналоги данных рассеяния для быстро
убывающих потенциалов разрабатывались \cite{53,39, 84}. Скалярные
спектральные задачи являются частными случаями матричных, а для них,
включая обобщения, показано, что потенциалы являются абелевыми
функциями \cite[стр.\,110]{29}, \cite[\S\,2]{25} и \cite{dik1,dik3}.
Добавим что способы вывода формул типа Матвеева--Итса  сейчас хорошо
известен \cite{4,5} и легко переносятся на другие уравнения.

\subsection{Конечнозонные операторы}
Пусть имеется изоспектральная деформация $\Psi_t=\widehat A\Psi$
задачи \reff{gd}. Когда эти уравнения могут быть проинтегрированы
явно? Если коэффициенты  $\widehat L$ зависят от обеих переменных
$(x,\,t)$, то первое уравнение $\widehat
L([u(x,t)],\,\partial_x)\,\Psi=\lambda\Psi$ представляет собой
фактически  задачу с двумя параметрами $\lambda$ и $t$, причем
зависимость от $t$ неизвестным образом скрыта в потенциале.
Простейший вариант --- считать $[u(x,t)]$ зависящим от стационарной
переменной $\xi=x-\alpha\,t$. Переписывая уравнения через переменную
$\xi$, получаем что время $t$ исчезает в первом операторе и у него
найдется решение вида $\Psi(x,\,t)=T(t)\,\Xi(\xi)$. Естественно
появившееся разделение переменных,  сразу же порождает параметр
разделения $\mu$, зависимость $T(t)=\exp(\mu\,t)$ и коммутирующий
оператор $\widehat A\Psi=\mu\Psi$. Мы будем использовать эту
простейшую мотивацию для известного \cite{29} определения
конечнозонного оператора. А именно, потенциал $\pot$ в линейном
дифференциальном выражении \reff{gd} называется {\em
конечнозонным\/} или {\em алгебро-геометрическим\/} если существует
парный дифференциальный оператор $\widehat A(\pot)$, такой что оба
оператора имеют общую собственную функцию $\Psi$ с соответствующими
собственными значениями $\widehat L\Psi=\lambda\Psi$ и $\widehat
A\Psi=\mu\Psi$. Cм. также \cite[стр.\,42]{2} и \cite{25,5}.

\section{$\Psi$-функция и алгебраические кривые}
\subsection{Алгебраические представления}

Если оператор (т.\,е. потенциал) конечнозонный, то $\Psi$-функция
выписывается явно в \pot-представлении. Для этого достаточно
последовательно исключить производные из уравнений $\widehat
L\Psi=\lambda\Psi$ и $\widehat A\Psi=\mu\Psi$ вплоть до формулы
$$
\Psi_x=G(\lambda,\,\mu;\,\pot)\,\Psi\,.
$$

{\bf Теорема 6.} {\em $\Psi$-функция для конечнозонных потенциалов
дается квадратурой вида\/}
\begin{equation}\label{exp}
\Psi(x;\lambda) = \exp\!\int\limits_{ x_{\sss 0}^{}}^{\,x} \!\!
G(\lambda,\,\mu;\,\pot)\,dx\,,
\end{equation}
{\em где $G$ --- рациональная функция по $(\lambda,\,\mu)$ с
коэффициентами, являющимися дифференциальными полиномами от $\pot$.
Соответствующая спектральная кривая имеет вид\/}
\begin{equation}\label{curve}
W(\mu,\lambda) \equiv
\mu^n+a_1^{}(\lambda)\,\mu^{n-1}+a_2^{}(\lambda)\,\mu^{n-2}+ \cdots
+a_n(\lambda)=0\,.
\end{equation}
Алгебраическое соотношение \reff{curve}, получается подстановкой
\reff{exp} в любой из операторов или, что тоже  самое, исключая
производные до конца. В выражении \reff{curve} величины
$a_k^{}(\lambda)$ являются полиномами по $\lambda$ и
дифференциальными полиномами от \pot. Таким образом вместо условия
разрешимости \reff{curve}, записываемого обычно в детерминантной
форме $\mbox{det}\big(\widetilde A(\lambda) - \mu\,I\big)=0$, где
$\widetilde A(\lambda)$ --- матричное представление оператора
$\widehat A$ в некотором базисе, предъявляется решение \reff{exp}, а
уравнение кривой \reff{curve} есть его следствие. Из формулы
\reff{exp} сразу следует, что группы Галуа общих конечнозонных
операторов устроены как и в случае уравнения \reff{schr}.

{\bf Предложение 2.} {\em Дифференциальная группа Галуа
$\mathrm{Gal}$ конечнозонного оператора \reff{gd} сопряжена группе
диагональных матриц вида
$\mathfrak{G}=\mathrm{Diag}(\alpha,\beta,\ldots,\gamma)
\subset\mathrm{SL}_n(\mathbb{C})$\/}.

Способы построения коммутирующего оператора $\widehat A$ используют
иерархии $[L,\,A]$-пар для интегрируемых нелинейных дифференциальных
уравнений, которые подробно прописаны в литературе. В силу первого
уравнения $\widehat L\Psi=\lambda\,\Psi$, второй оператор можно
считать полиномиальным пучком по $\lambda$. В таком виде процесс его
построения алгоритмизируется. Соотношения на коэффициенты
рекурсивны, но обозримый вид они имеют только для спектральных задач
порядка два. Характер и размер получающихся формул можно оценить по
работам \cite{68,45}.

\subsection{Неалгебраические (трансцендентные) представления}
Формулы (\ref{exp}--\ref{curve}) дают не\-яв\-ное описание решений
как функций от $\lambda$. С другой стороны мы можем перейти к
параметрически трансцендентному, но явному представлению
$W\big(\mu(\tau),\lambda(\tau)\big)=0$ и тогда обе спектральные
задачи могут рассматриваться как задачи со спектральным параметром
$\tau$, лежащем на алгебраической кривой \reff{curve}. Все они
интегрируются одновременно. Выбор алгебраического описания
(\ref{exp}--\ref{curve}) или трансцендентного, типа $\Psi(x;\tau)$,
является вопросом эффективизации и зависит от контекста задачи, хотя
явные формулы всегда предпочтительнее. В качестве иллюстрации,
рассмотрим пример, где все формулы выписываются до конца \cite{br}:
\begin{equation}
\label{burn} \mu^2=\lambda^5-\lambda\,, \qquad\qquad
\left\{\lambda=\frac{\vartheta_4(\tau)}{\vartheta_3(\tau)}\;,\qquad
\mu=i\,\frac{\vartheta_2^2(\tau)}{\vartheta_3^3(\tau)}\,
\vartheta_4(2\tau) \right\}\;.
\end{equation}
Здесь $\vartheta(\tau)=\theta(0|\tau)$ --- классические
$\vartheta$-константы Якоби.  Используя \reff{burn}, получаем задачу
$$
\Psi_{\mathit{xx}}-u(x)\,\Psi=
\frac{\vartheta_4(\tau)}{\vartheta_3(\tau)}\,\Psi\;.
$$
Кривая \reff{burn}\  обладает максимальной симметрией
(автоморфизмами) для кривых рода $g=2$ и поэтому реализуется как
накрытие эллиптической кривой
\begin{equation}\label{tor}
\wp'(\alpha)^2=4\,\wp(\alpha)^3-\frac53\,\wp(\alpha)\pm
\frac{27}{7}\,\sqrt{2\,}
\end{equation}
с модулем
$\varkappa=\frac{1+i\sqrt{2}}{2}$. Это ведет к тому, что все решения
выписываются в эллиптических функциях.  Формулы извлекаются из
теории эллиптических солитонов \cite{34,3} через известные там
анзацы. Соответствующая техника почти стандартизирована, поэтому
конкретные выражения для $u(x)$  через $\wp$-функции и
$\Psi\big(x;\alpha(\tau)\big)$ в форме Альфена--Эрмита мы не
приводим, ограничившись редукцией соответствующей $\Theta$-функции к
функциям Якоби:
$$
\Theta \Big(
\begin{smallmatrix}z_1^{}\\
\mathstrut z_2^{}\end{smallmatrix}\Big|
\ds\begin{smallmatrix}\varkappa&\frac12_{}\\\frac12\,&
\varkappa\end{smallmatrix}
\Big)=\mfrac12\,\big(\theta_3^{}\!(z_1^{}|\varkappa)
+\theta_4^{}\!(z_1^{}|\varkappa) \big)\,
\theta_3^{}\!\big(z_2^{}\big|\varkappa\big) +
\mfrac12\,\big(\theta_3^{}\!(z_1^{}|\varkappa)
-\theta_4^{}\!(z_1^{}|\varkappa)
\big)\,\theta_4^{}\!\big(z_2^{}\big|\varkappa\big)\;.
$$
Величина $\alpha$, как параметр накрываемого тора \reff{tor},
является промежуточным объектом и, поэтому, нам остается выписать
только явную формулу, связывающую $\alpha$ с глобальным параметром
$\tau$ на самой кривой \reff{burn} \cite{br}:
$$
-2\,\big(1\mp\sqrt{2\,}\big)\,\wp(\alpha)-\frac13\pm
\frac23\,\sqrt{2\,}= \frac{\vartheta_3^2(2\tau)}
{\vartheta_4(\tau)\,\vartheta_3(4\tau)}\,,
$$
Эту формулу можно рассматривать как трансцендентное (не
алгебраическое) представление  кривой \reff{burn}  мероморфными
функциями накрываемой и накрывающей поверхности. Дополнительные
детали почти без изменений обозначений переносятся сюда из \S\,4 в
\cite{br}.

Поскольку интегрируемые  задачи характеризуются тем, что
спектральным параметром является мероморфная функция $\lambda(\tau)$
на  кривой конечного рода, то коммутация пучков приводит, помимо
явной $\Psi$, к тому что спектральными параметрами для них может
быть взята одна величина $\tau$, лежащая на  общей римановой
поверхности (спектр задачи). Когда $\tau$ пробегает по всем точкам
спектра, $\Psi(x;\tau)$ пробегает по всем решениям этих спектральных
задач.

\section{Уравнения Дубровина}
\noindent В этом параграфе мы продемонстрируем общий рецепт вывода
уравнений Дубровина на величины $\gk$. Способ их вывода по Драшу
\cite{49_2} (см. также \cite[\S\,7]{31}, \cite{43,46}) использует
уравнение на ядро резольвенты и не переносится непосредственно на
произвольные спектральные задачи. В самом деле, для оператора
Штурма--Лиувилля ядро совпадает с квадратичной формой двух его
линейно независимых решений $\Psi_{\!1,2}^{}$ \cite[стр.\,83]{28}.
По нему строится спектральная матрица плотности \cite{14}, а вслед
за ней, и все что связано с разбиением единицы оператора на
бесконечном интервале \cite{15}. Для операторов высших порядков
обобщения в контексте спектральной теории далеко не очевидны. Тем не
менее в контексте интегрируемости по Лиувиллю, техника резольвент
известна \cite{dik4} и распространяется на некоторые операторные
пучки \cite{dik1,dik3}, хотя и с потерей эффективности для уравнений
на резольвенты (аналоги функции $R$ \cite{yal}).  Но как явствует из
\reff{exp}, необходимость состоит не в $R$, а в коэффициентах
фундаментальной мероморфной функции $G(\lambda,\mu;\pot)$.

\subsection{Уравнения 3-го порядка в иерархиях Гельфанда--Дикого}
Предлагаемые ниже выводы справедливы для произвольных конечнозонных
операторных пучков, но чтобы избежать излишней общности, мы
рассмотрим следующие спектральные уравнения:

\begin{equation}
\label{sp3} \Psi'''+u(x)\,\Psi' + v(x)\,\Psi=\lambda\,\Psi\,.
\end{equation}
Анзац для коммутирующего оператора имеет вид
\begin{equation}\label{ABC}
A(x;\lambda)\,\Psi''+B(x;\lambda)\,\Psi'+C(x;\lambda)\,\Psi
=\mu\,\Psi\,.
\end{equation}
Из \reff{ABC} и \reff{sp3} получаем уравнение кривой
(формула сокращена)
$$
\mu^3+(A''+3\,B'-2\,u\,A+3\,C)\,\mu^2+(\ldots)\,\mu+(\ldots)=0\,.
$$
Коэффициент перед  $\mu^2$, как константу $k(\lambda)$, можно
положить равным нулю, так как это соответствует сдвигу параметра
$\mu \to \mu-k(\lambda)/3$. В результате получаем
$$
C=-\mfrac13\,A''-B'+\mfrac23\,u\,A\,,
$$
\begin{equation}\label{curve3}
\mu^3+\mfrac13\, \big( A\,A^{\sss (\mathrm{IV})}
+(2\,A'+3\,B)\,A'''+ 3\,A\,B'''-A''^2 -u\,A\,A''-\cdots\big)\,\mu+
(\ldots)=0\,.
\end{equation}
$\Psi$-функция имеет вид
\begin{equation} \label{lnpsi}
\frac{\Psi'}{\Psi}=\frac{(A'+B)\,(\mu-C)-v\,A^2+A\,C'} {3\,A\,\mu +
A\,A''+u\,A^{2^{\mathstrut}}+3\,B\,(A'+B)}\,.
\end{equation}
Эта важная общая формула нам еще понадобится. Вывод построим так,
чтобы, чтобы методика без изменений переносилась на высшие порядки.

Необходимо определить всю совокупность нулей $\Psi$ на
алгебраической кривой, которые зависят от $\lambda$ и разбросаны по
листам римановой поверхности. Это всегда делается обращением в
полином рациональной по $\mu$ функции \reff{lnpsi} с помощью
уравнения кривой \reff{curve3}
$$
W(\mu,\,\lambda) \equiv \mu^3+Q(\lambda)\,\mu+T(\lambda)=0\,.
$$
В результате получаем формулу
\begin{equation}\label{psi3}
{\frac {\Psi'}{\Psi}}=-{\mfrac {3\,A\,\mu^2-(A\,A''+
u\,A^2+3\,B^2+3\,B\,A' )\,\mu +A^{\mbox{{\tiny({\sc\scriptsize
iv})}}} + \cdots}
{(3\,v-u')\,A^3+(3\,B\,u+2\,A'\,u+2\,A'''+3\,B'')\,A^2+
3\,(A''+B')\,B\,A+3\,B^2\,(A'+B)}}\,.
\end{equation}
Все нули $\Psi$ являются полюсами для $\Psi'\!/\Psi$  и поэтому
обозначим знаменатель \reff{psi3} как
\begin{equation}\label{pi}
\begin{array}{l}
R(\lambda;\,x)  = (3\,v-u')\,A^3+(3\,B\,u+2\,A'\,u+
2\,A'''+3\,B'')\,A^2+
3\,(A''+B')\,B\,A+3\,B^2\,(A'+B) =\\ \\
\phantom{R(\lambda;\,x)} =  a\!\cdot\!(\lambda-\gamma_1^{})\cdots
(\lambda-\gamma_g^{})\,,
\end{array}
\end{equation}
где $a$ числовая константа. Вторую координату нуля
$\mu_k^{}=\mu\big(\gk(x)\big)$, т.\,е. номер листа, определим как
корень кубического уравнения
\begin{equation}\label{uniform}
\mu_k^3+Q\big(\gk(x)\big)\,\mu_k^{} +T\big(\gk(x)\big)=0\,.
\end{equation}
Формулы (\ref{lnpsi}, \ref{psi3}) и \reff{pi} содержат
полную информацию для вывода всех дальнейших соотношений. Используя
\reff{lnpsi}
$$
\frac{\Psi'}{\Psi}= \frac{(A'+B)\,(\mu-C)+A\,C'-v\,A^2}{A\,\mu-D}\,,
$$
где введено обозначение
$$
-3\,D=A\,A''+u\,A^2+3\,B\,(A'+B)\,,
$$
получаем  важное соотношение --- формулу для второй координаты. А
именно, когда $\lambda \to \gk(x)$,
\begin{equation}\label{mu}
\mu_k^{} \to -\mfrac13\left(A''+u\,A+3\,\mfrac{B}{A}\,(A'+B)
\right)\,.
\end{equation}
Далее, ссылаясь на формулу \reff{mu} будем
считать ее равенством. Используя выражение для $Q(\lambda)$ как
функции от $\pot$ (формула \reff{curve3}), придадим формуле
\reff{psi3} следующий вид:
$$
\frac{\Psi'}{\Psi}= \frac{(A'+B)\,(\mu-C)+A\,C'-v\,A^2}{A\,\mu-D}=
-3\,\frac{(\mu^2+Q)\,A^2+A\,D\,\mu+D^2}{A\,R}+\frac{A'+B}{A}\;.
$$
Устремляя в последней формуле $\lambda \to \gk$ и используя
\reff{mu}, получаем искомые уравнения.

{\bf Теорема 7.}  {\em Координаты нулей $(\gk,\,\mu_k^{})$
$\Psi$-функции для конечнозонного спектрального пучка \reff{sp3}
удовлетворяют системе дифференциальных уравнений:
\begin{equation}\label{dubr}
\quad\gamma'_k=3\,A\big(\gk;\,\pot\big)\,\mfrac{3\,\mu_k^2+ Q(\gk)}
{\displaystyle a \prod _{j \ne k}(\gk-\gamma_j^{})}\,, \qquad \mu'_k
= -3\,A\big(\gk;\,\pot\big)\, \mfrac{Q'(\gk)\,\mu_k^{}+T'(\gk)}
{\displaystyle a \prod _{j \ne k}(\gk-\gamma_j^{})}\,.
\end{equation}
Вторая координата нулей $\mu_k^{}$ имеет вид}
$$
\mu_k^{} =\mfrac{D\big(\gk;\,\pot\big)}{A\big(\gk;\,\pot\big)}\,.
$$
Вторая группа уравнений \reff{dubr} следует из \reff{uniform}
взятием производной. Первая группа уравнений в \reff{dubr}
появлялась в работах \cite{44,45} в довольно громоздких обозначениях
описывающих стационарную иерархию уравнений Буссинеска. Мотивировки,
введение второй координаты $\mu_k^{}$ и доказательства  не
приведены.  Важное наблюдение авторов \cite{44} состоит в том, что
появившийся полином $3\,\mu^2+Q(\lambda)$ есть в точности
производная $W_{\!\mu}(\mu,\lambda)$. Этим видимо преследуется цель
(чего, однако, там не сделано) привести уравнения Дубровина
\reff{dubr} к задаче обращения Якоби с голоморфными дифференциалами
\begin{equation}\label{cheb}
d\,\omega = \frac{P(\mu,\lambda)}{W_{\!\mu}(\mu,\lambda)}\,
d\lambda\,.
\end{equation}
Мы вплотную подходим к интегрированию  уравнений \reff{dubr}  и
поэтому следует сделать важное замечание о характере этих уравнений.
В упомянутых работах это не обсуждается. Уравнения  \reff{dubr} не
автономны, так как содержат потенциал в $A$-функции и теряют смысл
пока не решен вопрос об представлении $A(\gk;\,\pot)$ через
$(\gamma,\,\mu)$. Добавим, что эффективные способы описания
конечнозонных потенциалов и их спектральных характеристик в  задачах
порядка выше 2 отсутствуют. Недавно пробел стал восполняться как
собственно для спектральных задач \cite{44, 45}, так и для
динамических систем \cite{75}.

\section{Временная динамика}
\noindent Динамика полюсов $\gk$ во времени определяется, как только
задано эволюционное уравнение:
\begin{equation}\label{time}
\quad\Psi_{\!t}= \widetilde
A(\lambda;\,x,\,t)\,\Psi_{\!\mathit{xx}}+ \widetilde
B(\lambda;\,x,\,t)\,\Psi_{\!x}+\widetilde
C(\lambda;\,x,\,t)\,\Psi\,,\qquad \widetilde C=-\mfrac13\,\widetilde
A_{\mathit{xx}}-\widetilde B_x+
\mfrac23\,u(\lambda;x,\,t)\,\widetilde A\,.
\end{equation}
В самом деле, кривая \reff{curve} не зависит от $t$ и формула
\reff{exp} справедлива в любой момент времени $t$. Таким образом
формула \reff{exp}, помимо выполнения  стандартного условия
$\Psi(x_{\sss 0})=1$, обеспечивает подвижную нормировку базиса, так,
что уравнение \reff{time} остается справедливым и для \reff{exp}.
Если установить или наложить свойство периодичности на потенциал
\pot, то \reff{exp} диагонализует матрицу монодромии и удовлетворяет
уравнению \reff{ABC} во все моменты времени. Подитожим  результаты
относительно дивизора  $(\gk,\,\mu_k^{})$ (см. также \cite{44,45}).

{\bf Теорема 8.} {\em  Пусть \potxt\  --- конечнозонный потенциал
спектральной задачи
$$
\Psi_{\!\mathit{xxx}}+u(x,\,t)\,\Psi_{\!x}+ v(x,\,t)\,\Psi
=\lambda\,\Psi
$$
с коммутирующим оператором
\begin{equation}\label{commut}
A(\lambda;\,x,\,t)\,\Psi_{\!\mathit{xx}}+
B(\lambda;\,x,\,t)\,\Psi_{\!x}+C(\lambda;\,x,\,t)\,\Psi=\mu\,\Psi\,,
\qquad
 C=-\mfrac13\,A_{\mathit{xx}}-B_{x}+ \mfrac23\,u\, A\,.
\end{equation}
и алгебраической кривой
$$
\mu^3+Q(\lambda)\,\mu+T(\lambda)=0\,.
$$
Пусть динамика $\Psi(\lambda;\,x,\,t)$ описывается уравнением
\reff{time}. Тогда

$\bullet$ Функции $\big(\gk(x,\,t),\,\mu_k^{}(x,\,t)\big)$
удовлетворяет системе дифференциальных уравнений
\begin{equation}\label{dubr1}
\left\{
\begin{array}{ccl}
\displaystyle\frac{d}{dx}\, \gk&=&
\displaystyle3\,A(\gk)\,\frac{3\,\mu_k^2+Q(\gk)}
{\displaystyle a\prod _{j \ne k}(\gk-\gamma_j^{})}\\ \\
\displaystyle\frac{d}{dx}\, \mu_k^{}&=&\displaystyle-3\,A(\gk)\,
\frac{Q'(\gk)\,\mu_k^{}+T'(\gk)} {\displaystyle a \prod _{j \ne
k}(\gk-\gamma_j^{})}
\end{array}
\right.\,,
\end{equation}
\begin{equation}\label{dubr2}
\left\{
\begin{array}{ccl}
\displaystyle\frac{d}{dt}\, \gk&=&3\, \Big\{A(\gk)\,\widetilde
B(\gk)- \widetilde A(\gk)\,B(\gk)\Big\}
\,\displaystyle\frac{3\,\mu_k^2+Q(\gk)}
{\displaystyle a \prod _{j \ne k_{{\mathstrut}}}(\gk-\gamma_j^{})}\\ \\
\displaystyle\frac{d}{dt}\, \mu_k^{}&=& 3\, \Big\{\widetilde
A(\gk)\,B(\gk)- A(\gk)\,\widetilde B(\gk)\Big\}\, \displaystyle
\frac{Q'(\gamma_k^{{}^{{\mathstrut}}})\,\mu_k^{}+T'(\gk)}
{\displaystyle a \prod _{j \ne k}(\gk-\gamma_j^{})}
\end{array}
\right.\,.
\end{equation}

$\bullet$  Вторая координата, однозначно определяемая из соотношения
$$
\mu_k^{}(x,\,t)=\frac{D(\gk;\,x,\,t)}{A(\gk;\,x,\,t)}\,,
$$
сводит  уравнения {\rm $($\ref{dubr1}--\ref{dubr2}$)$} к автономному
виду и порождает главную формулу следов\/}
\begin{equation}\label{trace}
\potxt=S(\gamma,\,\mu)\,.
\end{equation}
{\em Доказательство\/}. Уравнения \reff{dubr1} уже были доказаны в
\S\,5 (теорема 7), а формула следов \reff{trace} существует
поскольку потенциал является абелевой функцией
\cite{29,25,dik1,dik3}. Осталось доказать \reff{dubr2}. Подставляя
$\Psi_{\!\mathit{xx}}$ из уравнения \reff{commut} в уравнение
\reff{time}, получаем
$$
\frac{\Psi_{t}}{\Psi}= \frac{A\,\widetilde B-\widetilde A\,B}{A}\,
\frac{\Psi_{\!x}}{\Psi}+\widetilde C+(\mu-C)\,\frac{\widetilde
A}{A}\,.
$$
Осуществляя в данном выражении предел $\lambda \to \gk$, приходим к
системе уравнений гидродинамического типа со связью \reff{uniform}
$$
\dot{\!\gamma}_k=\frac{A\,\widetilde B-\widetilde A\,B}{A}\,
\gamma_k'\,,
$$
откуда,  в силу \reff{dubr1},  следуют уравнения \reff{dubr2}.
\hfill$\blacksquare$

{\bf Предложение 3.} {\em Формулы для тригональных конечнозонных
функций $\Psi(x;\lambda)$ типа \reff{final} получаются из формулы
\reff{exp}, используя \reff{dubr1}, переходом к интегрированию
абелевых дифференциалов 3-го рода на негиперэллиптической
тригональной кривой \reff{curve3}\/}.

Мы такие формулы здесь не приводим, поскольку для этого нам
потребовалось бы задействовать способы построения присоединенных
полиномов в \reff{cheb} \cite{80} и предъявить общие главные формулы
следов \reff{trace}. В некоторых частных случаях и род{\'а}х, это
нетрудно сделать используя результаты \S\,9.

При необходимости, <<спектральное>> описание задачи выводится после
перестановки пределов и параметров в упомянутых интегралах 3-го
рода. Естественным является и обратный переход. Прямые
<<спектральные формулировки>> для общих операторных пучков могут
быть весьма не очевидным в силу многих причин: гладкость,
вещественность и несингулярность потенциалов, нетривиальный способ
вхождения параметра $\lambda$ в пучок, определение функциональных
пространств, к которым принадлежит потенциал (задачи рассеяния) и
норм на этих пространствах, аналоги <<разложения единицы>> и
равенств Парсеваля, возможные спектральные особенности, матрицы
плотности  и т.\,д. Как следует из вышеописанных построений, точно
решаемые случаи предъявляются явно коль скоро известна
$\Psi$-функция. Для этого достаточно переставить параметры и пределы
в упомянутых интегралах и домножить $\Psi$-функцию на нормировочный
множитель. Тогда, например, все решения в элементарных аналитических
функциях будут соответствовать случаям когда кривая вырождается в
кривую рода ноль.

\section{Формулы следов}

\subsection{Рациональные формулы следов}
Главная формула следов \reff{trace} --- это фактически окончательный
ответ, а ее наличие, в виде  рациональной симметрической функции
$S$, является необходимым пунктом квадратурной интегрируемости. Без
этого, формула типа \reff{exp} почти не несет информации, так как
приложима к любым линейным дифференциальным уравнениям. Имея следы,
уравнения \reff{dubr} становятся дифференциальной формой
интегральной задачи обращения. Какими будут эти интегралы,
голоморфными, мероморфными или логарифмическими, для интегрируемости
не имеет значения, так как появляется полный набор констант
интегрирования. Логарифмические интегралы возникают если имеются
кратные точки ветвления. Это теория солитонов в виде экспонент или
солитоны на фоне конечнозонных решений. Мероморфные интегралы
возникают в современных задах с нелинейной эволюцией на якобианах,
которые известны как <<бильярды>>.

Появление второй координаты $\mu_k^{}$ естественно, так как
произвольная абелева функция  есть симметрическая комбинация верхних
пределов абелевых интегралов в задаче обращения Якоби и обе
координаты становятся равноправными.  Дальнейшие примеры см.  в
\S\,9. При интегрировании гамильтоновых систем методом разделения
переменных, этот факт  стал разрабатываться в 90-х годах (см.
например \cite[\S\,3]{kuz}) и проявляется в том, что переход  от
координат/импульсов $(p,\,q)$ (или переменных Остроградского
$u,\,u_x,\ldots$) к переменным разделения $(\gamma,\mu)$ содержит
вообще говоря оба набора функций и представляет собой расширение
координатного метода разделения переменных.

Следует отметить, что фундаментальная роль различных версий формул
следов указывалась Матвеевым еще в 1975 г. \cite{matv,29}. Причем
нелинейные и динамические варианты таких формул таковы, что при
рассмотрении уравнения Кортевега--де Вриза, уравнение Шредингера как
таковое можно даже не рассматривать \cite{matv}. Как станет видно из
примеров \S\,9 и следующего пункта, в негиперэллиптических ситуациях
имеет место  <<перепутывание>>  разновидностей <<следов>> и
переменных $(\gamma,\mu)$, но существует формула \reff{trace}, из
которой следуют остальные. Ее можно получать не обязательно  в виде
рациональной  функции.

\subsection{Алгебраические и трансцендентные формулы следов}

Содержание этого пункта основано на одном наблюдении, взятом из
анализа примеров в \S\,9. Возникающие там полиномиальные тождества
говорят о том, что возможно получать формулы для потенциалов в
алгебраической и трансцендентной, а не рациональной симметрической
форме. В самом деле, полиномы (см. \S\,9), содержащие необходимую
информацию, влекут варианты формул, которые, с технической точки
зрения, легче выводятся  и появляются первично. Так для {\sc Примера
5} на стр.\,\pageref{5}, анализируя набор интегралов Новикова
$E_k(\pot)$, получаем следующее выражение:
\begin{equation}\label{3}
u^3+3\,\alpha\,u+3\,E_2^{}+27\,\sum\limits_{k=1}^{4}\,\gamma_k^2(x)=0\,.
\end{equation}
Рассматривая его как полином по $u$, находим корни, которые
тем не менее являются аналитическими однозначными функциями от $x$.
Для этого воспользуемся эллиптическими представлениями для решения
алгебраических уравнений 3--4-й степени:
\begin{equation}\label{trans}
u=\wp\big(\omega(x);
-12\alpha,-12E_2-108\textstyle\sum\gamma_k^2(x)\big)\;,
\end{equation}
где $\omega(x)$ --- любой полупериод, соответствующий инвариантам
данной функции Вейерштрасса, а величины $\gk(x)$ определяются в
предложении 5 (\S\,9). Переходя в формуле \reff{trans} к явным
$\vartheta$-представлениям, после некоторых упрощений, получаем
ответ:
\begin{equation}\label{transs}
u=-\frac{3}{2\alpha}\,\left\{E_2+9\sum\limits_{k=1}^{4}
\gamma_k^2(x)\right\}
\frac{\vartheta_2^8(\tau)+\vartheta_3^8(\tau)+\vartheta_4^8(\tau)}
{\big(\vartheta_2^4(\tau)+\vartheta_3^4(\tau)\big)
\big(\vartheta_3^4(\tau)+\vartheta_4^4(\tau)\big)}\,,
\end{equation}
где эллиптический модуль $\tau=\tau(x)$ определяется, используя
модулярный инвариант Клейна $J(\tau)$, из решения трансцендентного
уравнения\footnote{Здесь уместно заметить, что решение эллиптической
модулярной задачи обращения (оно используется при выводе формулы
\reff{transs}), изложенное в книге Н.Ахиезера {\em ``Элементы теории
эллиптических функций''} (1970) на стр.\,51 или в 3-м томе {\em
Матем.\,Энц\/}. на стр.\,789, является не верным (не связано с
опечаткой).}
$$
J(\tau)=\frac{4\alpha^3}{4\alpha^3-9E_2-81\sum\gamma_k^2(x)}\,.
$$
Решение таких уравнений, при заданной правой части, как известно,
единственно. Формулы (\ref{trans}--\ref{transs}) доставляют примеры
{\em симметрических трансцендентных следов\/}. Случай когда
$\alpha=0$ является вырожденным (см. {\sc Пример 1}) и требует
отдельных формул. Они несколько проще, но мы их не приводим.
Подобным способом можно порождать много примеров абелевых функций
являющихся радикалами или алгебраическими функциями от других
абелевых функций. Например, как это следует из  формулы \reff{4},
получаем однозначное представление
$$
18\,
\sqrt[3]{\mathstrut\gamma_1^{}\,\gamma_2^{}\,\gamma_3^{}\,\gamma_4^{}}
=u''(x)+2\,u^2(x)+2\,\alpha\,.
$$

Общий механизм получения <<трансцендентных следов>> состоит в
исключении производных от потенциала из полного набора соотношений,
определяющих интегралы $E_k(\pot)$ и переменные разделения $\gk$.
Решая возникающие алгебраические уравнения в однозначных функциях,
получаем ответ. Эти уравнения  (возможно даже их порядки) не
единственны. Переходы между ними осуществляют различные
$(\gamma,\mu)$-представления абелевых функций.  Аналогичные формулы
выписываются для других уравнений, включая гиперэллиптический случай
КдВ. Здесь мы ограничимся лишь примером (\ref{trans}--\ref{transs})
и приведенными комментариями, поскольку в полной мере эта тема
затрагивает вопросы униформизации алгебраических кривых и
трансцендентных тождеств между $\Theta$-функциями, причем
различающихся родов.

\subsection{Трансцендентные соотношения между $\Theta$-функциями}

Обратное  проявление описанного выше представляет не меньший
интерес. В самом деле, сопоставляя формулы \reff{transs} и
\reff{trans2}, будем иметь
\begin{equation}\label{transAbel}
-\frac29\,\alpha\,\sum\limits_{k=1}^{4}\,\frac{\gamma_k^2\,\mu_k^{}}
{\displaystyle\prod\limits_{j \ne k}\,(\gk-\gamma_j^{})} =
\left\{E_2+9\sum\limits_{k=1}^{4} \gamma_k^2(x)\right\}
\frac{\vartheta_2^8(\tau)+\vartheta_3^8(\tau)+\vartheta_4^8(\tau)}
{\big(\vartheta_2^4(\tau)+\vartheta_3^4(\tau)\big)
\big(\vartheta_3^4(\tau)+\vartheta_4^4(\tau)\big)}\,.
\end{equation}
Записывая теперь в этой формуле
$\Theta(x\boldsymbol{U}+\boldsymbol{D})$-пред\-став\-ле\-ния для
имеющихся там абелевых функций, получим  трансцендентное
$\Theta$-тождество.  В данном примере у нас возникло соотношение
между $\Theta$-функциями рода 4 для кривой \reff{curvekk}  через
эллиптические $\vartheta$-константы. Это имело место в силу того,
что появлялось алгебраическое уравнение 3-й степени \reff{3}. В
случае появления уравнений высших порядков мы либо вновь будем
приходить к эллиптическим $\vartheta$-константам\footnote{Модулярные
уравнения и, в частности, знаменитое решение уравнения 5-й степени
Эрмита.}, либо к представлению корней уравнений в $\Theta$-функциях
высших родов.

Близкая связь между  примерами классической лиувиллевской
интегрируемости и неалгебраической интегрируемости в модулярных
функциях не раз отмечалась в литературе. См. например работы
\cite{tah,hitchin} и библиографию в них. В работе \cite{brtheta}, в
виде формульных следствий теоремы 4, эта связь комментируется
подробнее. В частности, используя приведенные там результаты, можно
напрямую проверить соотношения (\ref{transs}--\ref{transAbel})
непосредственным дифференцированием с использованием уравнений
Дубровина в {\sc Примере 5}.

\section{Спектральные характеристики потенциалов}
\noindent Здесь мы покажем как получать спектральные характеристики
потенциалов для тригональных уравнений \reff{sp3}. Уровень описания
конечнозонных потенциалов этих задач далек от того, какой сейчас
имеется для уравнений типа КдФ. Только недавно этот пробел стал
восполняться \cite{36,61,45}. Предлагаемые в этом параграфе формулы
являются обобщениями результата, который впервые был получен в
работе \cite{16} для уравнения КдВ.

{\bf Предложение 4.} {\em Пусть $u(x)$ --- произвольный $2$-зонный
потенциал  уравнения \reff{schr}. Тогда $\Psi$-функция дается
выражением
$$
\Psi(x;\lambda)=\sqrt{R\,}\,\exp\!\int\!\frac{\mu}{R}\,dx\,, \qquad
R=\lambda^2+\left(c_1^{}-\mfrac{u}{2}\right)
\lambda+\mfrac{3}{8}\,u^2-
\mfrac{1}{8}\,u_{\mathit{xx}}-\mfrac12\,\,c_1^{}\,u+c_2^{}\,.
$$
При  нормировке $x_{\sss 0} = a_{\sss 0} = 0$ для двух возможных
разложений потенциала
$$
\begin{array}{l}
u  =  \mfrac{6}{x^2_{{}_{\mathstrut}}} - \mfrac{4}{35}\,c_2^{}\,
x^2+ b\,x^4+c\,x^6+d\,x^8+
\cdots, \quad c_1^{}=0\,, \qquad \mbox{\rm\small
(6 свободных параметров \cite{16})}\\
u  = \mfrac{2^{{}^{\mathstrut}}}{x^2} -
\mfrac45\,c_2^{}\,x^2+b\,x^3+c\,x^4-
\mfrac{3}{10}\,c_1^{}\,b\,x^5+d\,x^6+\cdots,\quad \;\;
\mbox{\rm\small (7 свободных параметров)}
\end{array}
$$
соответствующие алгебраические кривые имеют вид\/$:$}
$$
\begin{array}{rl}
\mu^2 \!\!\!\!& =  \lambda^5 + 2\,c_2^{}\,\lambda^3+\mfrac {63}{2}
\,b\,\lambda^2 + \left
(\mfrac{162}{175}\,c_2^2+\mfrac{297}{4}\,c\right)_{{}_{\mathstrut}}
\!\lambda + \mfrac {1377}{35}\,b\,c_2^{}+\mfrac {1287}{2}\,d\,,\\
\mu^2 \!\!\!\!& =  \displaystyle  \lambda^5+2\,c_1^{}\,\lambda^4+
\left(c_1^2+2\,c_2^{}\right)^{{}^{\mathstrut}} \!\lambda^3+
\left(4\,c_1^{}\,c_2^{}-\mfrac{7}{2}\,c\right)_{{}_{\mathstrut}}
\!\lambda^2+{}
\\& \ds\phantom{=}
+\left( 2\,c_2^{}\,c_1^2-7\,c\,c_1^{}+\mfrac{54}{25}\,c_2^2-
\mfrac{81}{4}\,d\right )^{{}^{\mathstrut}} \!\lambda  -
\mfrac72\,c\,c_1^2+ \left(\mfrac{54}{25}\,c_2^2-
\mfrac{81}{4}\,d\right)\! c_1^{}+\mfrac {81}{64}\,b^2\,.
\end{array}
$$
Укажем общий способ получения таких формул. Всякий конечнозонный
потенциал мероморфен по $x$, поэтому, независимо от числа его
полюсов, помещая любой из них в $x=0$ и подставляя анзац
\begin{equation}\label{ans}
\pot=\frac{a_{-n}^{}}{x^n}+
\frac{a_{-n+1}^{}}{x^{n-1}}+ \cdots+ a_0^{}+a_1^{}\,x+\cdots
\end{equation}
в уравнение кривой \reff{curve} и группируя полярную часть по
степеням $x$ получаем. 1) Порядок $n$ и возможные значения
коэффициента $a_{-n}$. 2) Значения некоторых из коэффициентов $a_k$.
Число оставшихся из них равно числу констант интегрирования
стационарного уравнения. 3) Значения всех констант $\alpha$, $c_k$
через коэффициенты $a_k$. Они всегда определятся, так как уравнения
на них линейны. После разрешения пунктов 1--3), следующий, свободный
коэффициент разложения даст уравнение кривой \reff{curve}. Данный
способ определения коэффициентов $a_k$ конечно равноценен прямой
подстановке \reff{ans} непосредственно в стационарное уравнение, но
здесь мы автоматически определяем константы $c_k$, которые входят в
$\Psi$-формулу.

{\sc Пример 1.} Спектральная задача
\begin{equation}\label{spectrkk}
\Psi'''+u\,\Psi'+\mfrac12\,u'\,\Psi=\lambda\,\Psi\,.
\end{equation}
Коммутирующий оператор построим по стационарному
уравнению Каупа--Купершмидта \cite{64}
\begin{equation}\label{kk}
u_t^{}=u_{\mathit{xxxxx}}+5\,u\,u_{\mathit{xxx}}+
\mfrac{25}{2}\,u_x\,u_{\mathit{xx}} +5\,u^2\,u_x\,.
\end{equation}
Переходя к стационарной переменной $x \to
x-\alpha\,t$, получаем:
$$
-9\,\lambda\,\Psi''+\left(u^2+\mfrac12\,u''+\alpha\right)\Psi'-
\left(6\,u\,\lambda+2\,u\,u''+\mfrac12\,u'''\right)
\Psi=\mu\,\Psi\,.
$$
Здесь мы встречаемся с тем  фактом, что в алгебраическую кривую
$$
\mu^3+\Big(27\,\alpha\,{\lambda}^{2}-\mfrac92\,K\,\lambda+
E_1\Big)\,\mu+ 729\,{\lambda}^{5}+E_2\,\lambda^3+
\mfrac{27}{2}\,u\,K\,\lambda^2+
E_3\,\lambda+\mfrac{1}{8}\,(u''+2\,u^2+2\,\alpha)^2\,K=0
$$
попадают не только интегралы стационарного уравнения \reff{kk},
но и оно само:
$$
K \equiv \alpha\,u'+u^{\sss (\mathrm{V})}
+5\,u\,u'''+\mfrac{25}{2}\,u'\,u''+5\,u^2\,u'\,.
$$
$K$ должно быть нулем. Но этому не противоречит и выражение
$u''+2\,u^2+2\,\alpha=\mathit{const}$.
Если константа ненулевая, то получаем тривиальный
результат $u=\mbox{\it const}$. В противном случае имеем
кривую в вырожденной ситуации
$$
\mu^3+27\,\alpha\,\lambda^2\,\mu+729\,\lambda^5- 27\left
(\mfrac{3}{4}\,u'^{2}+u^3+3\,\alpha \,u \right)\lambda^3=0\,, \qquad
\mbox{род \ $g=1$.}
$$
Соответствующее решение легко вычисляется и имеет вид
$u=-3\,\wp\big(x+\mfrac34\,g_2^{}\,t\big)$. Для невырожденного
случая кривая \reff{curve} приобретает вид:
\begin{equation}\label{curvekk}
\mu^3+(27\,\alpha\,\lambda^2+E_1^{})\,\mu
+729\,\lambda^5+81\,E_2^{}\,\lambda^3+ E_3^{}\,\lambda=0\,,
\qquad
\mbox{ род \  $g=4$}\,.
\end{equation}
Нестандартность уравнения
\reff{kk} уже не раз отмечалась в теории солитонов. Обширные ссылки
приведены в \cite{22}, а теоремы суперпозиции для него были получены
совсем недавно  \cite{76}. Одно из двух возможных разложений
потенциала и кривая имеют вид
$$
u=-\frac{3}{x^2}+a\,x+b\,x^2+c\,x^3+d\,x^4+e\,x^5+\cdots, \qquad
\alpha=5\,b\,,
$$
$$
\begin{array}{l}
\mu^3+3\,(45\,b\,\lambda^2-52\,a^2\,b-128
\,a\,e-12\,c^2 )\,\mu+{}\\ \\
\quad\,{} +{\mfrac {9^{{}^{\mathstrut}}}{4}}\,\Big(324\,\lambda^4
-27\,(15\,a^2+28\,d)\,\lambda^2+116\,a^4+
336\,a^2\,d+384\,a\,b\,c+1536\,c\,e \Big)\,\lambda=0\,.
\end{array}
$$
Следующее разложение --- четная функция:
$$
u=-\frac{24}{x^2}+a\,x^2+b\,x^4-{ \mfrac
{a^2}{72}}\,x^6+c\,x^8+d\,x^{10}+\cdots, \qquad \alpha=110\,a\,.
$$
Выражение для кривой мы не выписываем.

{\sc Пример 2.} Спектральная задача
\begin{equation}\label{spectrsk}
\Psi'''+u\,\Psi'=\lambda\,\Psi
\end{equation}
обычно связывается с
уравнением Савады--Котера \cite{65}
\begin{equation}\label{sk}
u_t^{}=u_{\mathit{xxxxx}} + 5\,u\,u_{\mathit{xxx}} +5\,u_x\,
u_{\mathit{xx}}+5\,u^2\,u_x\,.
\end{equation}
Одно из двух возможных разложений стационарных решений \reff{sk}
имеет вид
$$
u=-\frac{12}{x^2}+a\,x^2+b\,x^3+c\,x^4-\left(
\mfrac{a^2}{36}\,x^6+\mfrac{a\,b}{22}\,
x^7+\mfrac{b^2+a\,c}{44}\,x^8+\mfrac {5\,b\,c}{156}\,x^9\right)
+d\,x^{10}+\cdots \,\,\, (\alpha=20\,a)\,.
$$
Отличительным является более сложная зависимость  от параметров. Четыре
коэффициента подряд при 6--9-й степенях $x$ не свободны, а определяются
предыдущими. Кривая имеет вид
$$
\mu^3+36\,(15\,a\,\lambda^2-49\,b^2)\,\mu+9\,
\lambda\,(3^4\,\lambda^4-3024\,c\,\lambda^2+1568\,a^3 - 2^6\,
21^2\,c^2 - 2^8 \,3^3 \, 637\,d) = 0\,.
$$
Аналогичные формулы получаются для второго возможного разложения
$$
u=-\frac{6}{x^2}+a+b\,x-\Big(\mfrac{a^2}{6}\,x^2-\mfrac{a\,b}{3}\,x^3\Big)
+c\,x^4+\Big(\mfrac{a^2\,b}{24}\,
x^5 -\mfrac {a^4-18\,a\,b^2}{648}\,x^6+
{\mfrac {b^3-a^3\,b}{252}}\,{x}^7\Big)+d\,x^8+ \cdots
$$
Не лишним будет привести пример неэллиптического конечнозонного потенциала,
но выражающегося в эллиптических функциях. Соответствующее решение стационарного
уравнения \reff{sk} было получено Шази \cite[стр.\,380]{66}
(Шази имел дело
с обыкновенным дифференциальным уравнением, которое он изучал
в контексте Пенлеве--анализа):
$$
u=-6\,\wp_1^{}-6\,\wp_2^{}\,, \qquad \wp_1^{} \equiv
\wp(x+12\,g_2^{}\,t-\Omega;\,g_2^{},g_3^{})\,, \quad \wp_2^{} \equiv
\wp(x+12\,g_2^{}\,t- \widetilde\Omega;\,g_2^{},\widetilde g_3^{})\,.
$$
Его спектральные характеристики не могут быть получены в  рамках
теории эллиптических солитонов. Кривая рода $g=4$, $\Psi$-функция и
аналог $R$-функции имеют вид \cite{b1}:
$$
\Psi(x;\lambda)=\exp\!\int\!\!
\mfrac {4\,(\wp_1^{}-\wp_2^{})^{2}\,\mu+9\,\lambda\,(8
\,\wp'_1\, \wp'_2+4\,(\wp_1^{}+\wp_2^{})\,(4\,\wp_1^{}\,\wp_2^{}- g_2^{})-
\lambda^2-4\,g_3^{}-4\,\widetilde g_3^{})}
{(2\,\wp'_1+2\,\wp'_2+\lambda)\,\mu+72\,\lambda\,(\wp'_1\,\wp_2^{}+
 \wp'_2\,\wp_1^{})+72\,(g_3^{}-\widetilde g_3^{})\,(\wp_1^{}-\wp_2^{})
+18\,(\wp_1^{}+\wp_2^{})\,\lambda^2}\,dx\,,
$$
\smallskip
$$
\mu^3-324\,g_2^{}\,\lambda^2\,\mu+729\,\lambda\,
\big((\lambda^2+4\,g_3^{}+4\,\widetilde g_3^{})^2-64\,g_3^{}\,
\widetilde g_3^{}\big)=0\,,
$$
\begin{equation}\label{R3}
R(x;\lambda)\equiv \Psi_1^{} \Psi_2^{} \Psi_3^{}=
(\lambda^2+4\,\wp_1'\,\lambda+4\,\widetilde g_3^{}-4\,g_3^{})
\,(\lambda^2+4\,\wp_2'\,\lambda-4\,\widetilde g_3^{}+4\,g_3^{})\,.
\end{equation}

{\sc Пример 3.} Является ли предыдущий потенциал
$u=-6\,\wp_1^{}-6\,\wp_2^{}$ со всеми произвольными инвариантами
$g_2^{}, \,g_3^{},\,\widetilde g_2^{},\, \widetilde g_3^{}$
конечнозонным для спектральной задачи \reff{spectrsk}\,? Ответ
положительный и этот пример, как нетривиальный в контексте уравнений
Дубровина, будет рассмотрен в \S\,9 ({\sc Пример 8}). Коммутирующий
операторный пучок выводится на основании вышеизложенных рассуждений
из следующего члена иерархии: уравнения 7-го порядка
\cite[стр.\,193, с учетом опечатки]{47}:
$$
u_t^{}=u_{7x}^{} + 7\left( u\,u_{5x}^{}+2\,u_x\,u_{\mathit{xxxx}}+
3\,u_{\mathit{xx}}u_{\mathit{xxx}}
+2\,u^2\,u_{\mathit{xxx}}+6\,u\,u_x\,u_{\mathit{xx}}+
u_x{}^{\!\!\!3}+\mfrac43\,u^3\,u_x\right)\,.
$$

\section{Формулы следов. Примеры}
\noindent
Как уже отмечалось, тригональные кривые проявляют б\'ольшую нестандартность и
трудность при классификации. В большей степени это проявляется
в тождествах следов. Следующий пример разберем подробно, чтобы
продемонстрировать характер доказательств.

{\sc Пример 4.} Задача \reff{spectrsk}. Коммутирующий пучок для рода
$g=4$ имеет вид:
$$
(-9\,\lambda+3\,u')\,\Psi''+(u^2-u''+\alpha)\,\Psi'-6\,\lambda\,u\,\Psi=
\mu\,\Psi\,.
$$
Получаем
$$
A(\gk;\,\pot)=-9\,\gk+3\,u'
$$
и уравнения \reff{dubr}  имеют неавтономный вид. Алгебраическая
кривая имеет вид \reff{curvekk}. Нам потребуется выражение для
интеграла $E_2$ в ней:
\begin{equation}\label{tmp1} E_2=u^{\sss
(\mathrm{IV})}+5\,u\,u''+\mfrac53\,u^3+\alpha\,u\,.
\end{equation}
Факторизованная форма полинома $R$ всегда выступает как определение
переменных $\gk$ \cite{br4}:
$$
3^{-7}\,R(\lambda;\,x)=\lambda^4-\mfrac23\,u'\,\lambda^3+
\mfrac19\,\big(u^{\sss (\mathrm{IV})}   + 3\,u\,u'' +2\,u'^2+
u^3+\alpha\,u\big)\,\lambda^2 +(\ldots)\,\lambda +(\ldots)\,,
$$
В итоге получаем формулу
\begin{equation}\label{tmp2}
u'=\mfrac23\,\sum\limits_{j=1}^{4}\,\gamma_j^{}\,.
\end{equation}
Воспользуемся вторым тождеством следов
$$
\mfrac19\,\big(u^{\sss (\mathrm{IV})} + 3\,u\,u'' +
2\,u'^2+u^3+\alpha\,u\big)=
\sum\limits_{k>j}^{4}\,\gamma_j^{}\,\gk\,,
$$
чтобы исключить из него и \reff{tmp1} $u^3$. В результате получаем
линейное соотношение на $u$ --- первый вариант главной формулы
следов:
\begin{equation}\label{trace1}
u =  -\mfrac{3}{2\,\alpha} \,\Big\{E_2
+\sum_{k=1}^{4}\,\big(\gamma_k'''+\mfrac{15}{2}\,
\gamma_k^2\big)\Big\}\,.
\end{equation}
Формально, \reff{trace1} уже отвечает положительно на вопрос о
формуле \reff{trace}, так как $\gamma',\,\gamma'',\,\gamma'''$
выражаются через $(\gamma,\,\mu)$ в силу  \reff{dubr} и \reff{tmp2}.
Но в этом примере это не реализуемо по причине непомерных
промежуточных формул. Простое решение получается после анализа
дифференциальных полиномов $Q(\lambda),\, T(\lambda)$. Используя
интеграл $E_3=u'''\,(u''-u^2-\alpha)\,u'+\cdots$, получаем
$$
3^{-7}\,R(\lambda;\,x)=\lambda^4-\mfrac23\,u'\,\lambda^3-
\mfrac{1}{27}\,(6\,u\,u'-6\,{u'}^2+2\,u^3-3\,E_1)\,\lambda^2 +
(\ldots)\,\lambda + \frac{6\,E_1\,u-E_3}{729}\,.
$$
Последнее слагаемое в этом выражении влечет интересный вариант
главной формулы следов
\begin{equation}\label{strange}
u=\frac{E_3+3^6\,\gamma_1^{}\,\gamma_2^{}\,
\gamma_3^{}\,\gamma_4^{}}{6\,E_1}\,.
\end{equation}
Другое, более
систематическое, решение дает анализ самих уравнений \reff{dubr},
которые с учетом \reff{tmp2},  мы запишем в виде
\begin{equation}\label{tmp3}
\frac{2\cdot 81\,d\,\gk}{3\,\mu_k^{2}+Q(\gk)}= \frac{\displaystyle
\sum\limits^{4}\, \gamma_j^{}-2\,\gk} {\displaystyle
\prod\limits_{j\ne k} \, (\gk-\gamma_j^{})}\,dx\,.
\end{equation}
Имеет место следующее

{\bf Предложение 5.} {\em Уравнения \reff{tmp3} эквивалентны задаче
обращения Якоби для кривой \reff{curvekk}
$$
\left\{
\begin{array}{ll}
\displaystyle\sum_{k=1}^4 \!\!\int\limits_{}^{(\gk,\,\mu_k^{})}
\!\!\!\!\!\frac{d\lambda}
{3\,\mu^2+Q(\lambda)}=a_1^{},&\displaystyle\quad
\sum_{k=1}^4 \!\!\int\limits_{}^{(\gk,\,\mu_k^{})} \!\!\!\!\!
\frac{\lambda\,d\lambda}
{3\,\mu^2+Q(\lambda)}=a_2^{},
\\ \\
\displaystyle \sum_{k=1}^4 \!\!\int\limits^{(\gk,\,\mu_k^{})}
\!\!\!\!\! \frac{\mu\,d\lambda} {3\,\mu^2+Q(\lambda)}=a_3^{}, &
\displaystyle\quad \sum_{k=1}^4 \!\!\int\limits^{(\gk,\,\mu_k^{})}
\!\!\!\!\!\frac{\lambda^2\,d\lambda}
{3\,\mu^2+Q(\lambda)}=a_4^{}-\mfrac{1}{81}\,x\,,
\end{array}
\right.
$$
где $Q(\lambda)=27\,\alpha\,\lambda^2+E_1$. Главная формула следов
имеет вид\/}:
\begin{equation}\label{trace2}
u = \frac16\,\sum_{k=1}^4
\,\gk\,\mu_k^{}\,\frac{\sum\limits^4\gamma_j^{} -2\,\gk}
{\displaystyle\prod _{j \ne k}(\gk-\gamma_j^{})}\,.
\end{equation}
{\em Доказательство\/}: 1, 2 и 4-е соотношения проверяются
непосредственным домножением правых частей \reff{tmp3} на
$(1,\,\gk,\,\gamma_k^2)$ соответственно, с последующим суммированием
и интегрированием. Для доказательства 3-го равенства воспользуемся
формулой для второй координаты \reff{mu}:
\begin{equation}\label{muk}
\mu_k^{}= \frac {27\,u\,\gamma_k^2-9\,(u'''+2\,u\,u'
)\,\gk+3\,u'''\,u'-2\,{u''}^2+
(\alpha+u^2)\,(u''+u^2+\alpha)+3\,u\,u'^2}{-9\,\gk+3\,u'}\,.
\end{equation}
Подставляя ее в формулу
$$
\sum\limits_{k=1}^4\,\frac{\mu_k^{}\,d\gk}
{3\,\mu_k^{}+Q(\gk)}
$$
убеждаемся, что это равно выражению
$$
3\,\frac{(2\,u''+u^2+\alpha)\,(u''-u^2-\alpha)\,\big(2\,u'-3\,\sum\,\gk\big)}
{\displaystyle \prod\limits_{k=1}^4\,(3\,\gk-u')}\,,
$$
которое в силу \reff{tmp2} равно нулю. Домножая правую часть
\reff{tmp3} на $\gk\,\mu_k^{}$, суммируя  и используя \reff{muk}
получаем формулу \reff{trace2}. \hfill$\blacksquare$

Сравнение формул   \reff{strange} и \reff{trace2} дает два представления
одной абелевой функции. В этом нетрудно усмотреть
проявление семейства $\Theta$-тождеств для
данной тригональной кривой.

{\sc Пример 5.} \label{5} Спектральная задача \reff{spectrkk}. Здесь
кривая имеет вид \reff{curvekk}, почти все формулы выглядят проще,
за исключением формулы следов типа \reff{trace1}, а уравнения
\reff{dubr} уже имеют автономный вид. В самом деле, факторизуя
полином $R(\lambda;\,x)$:
\begin{equation}\label{4}
\begin{array}{ccl}
3^{-7}\,R(\lambda;\,x)&=&\lambda^4-\mfrac16\,u'\,\lambda^3+
\mfrac{1}{18_{{\mathstrut}}}\,\big(u^{\sss (\mathrm{IV})}
 + 5\,u\,u'' +4\,u'^2+
2\,u^3+2\,\alpha\,u\big)\,\lambda^2 -\\ \\
&& {}-\mfrac{1}{324}\,(u'''+4\,u\,u')\, (u''+2\,u^2+2\,\alpha )\,
\lambda +\mfrac{1}{3^6
8}\,(u''+2\,u^2+2\,\alpha)^{3^{{\mathstrut}}}\,.
\end{array}
\end{equation}
получаем следующие тождества
$$
u'=6\,\sum\limits_{k=1}^{4}\,\gk, \qquad \mfrac{1}{18}\,\big(u^{\sss
(\mathrm{IV})} + 5\,u\,u'' +4\,u'^2+2\,u^3+2\,\alpha\,u\big)=
\sum\limits_{k>j}^{4}\,\gamma_j^{}\,\gk\,.
$$
Действуя также как и в предыдущем примере получаем аналог \reff{trace1}:
$$
u=\frac32\,\frac{45\,\big(\sum \gk \big)^2+ \sum
\big(2\,\gamma_k'''- 15\,\gamma_k^2\big) -2\,E_2}
{2\,\alpha-15\,\sum\gamma_k'}\,.
$$
Вторая координата нуля определяется выражением
$$
\mu_k^{}=3\,u\,\gk+\frac{(u''+2\,u^2+2\,\alpha)^2}{36\,\gk}\,,
$$
интегральная форма уравнений Дубровина имеет такой же вид как и в
предложении 4, а главная формула следов дается выражением
\begin{equation}\label{trans2}
u=\frac13\,\sum\limits_{k=1}^{4}\,\frac{\gamma_k^2\,\mu_k^{}}
{\displaystyle\prod\limits_{j \ne k}\,(\gk-\gamma_j^{})}\,.
\end{equation}

{\sc Пример 6.} Спектральная задача
$$
\Psi'''+u\,\Psi'+v\,\Psi=\lambda\,\Psi\,, \qquad
\Psi''+\alpha\,\Psi'+\mfrac23\,u\,\Psi=\mu\,\Psi\,, \qquad
\mbox{род\ \ $g=1$}\,.
$$
Пример проще предыдущих, но показателен, так как содержит
два потенциала $u,\,v$. Он соответствует
стационарным решениям системы уравнений
\begin{equation}\label{bsq}
v_t-v_{\mathit{xx}}+\mfrac23\,(u\,u_x+u_{\mathit{xxx}})=0, \qquad
u_t-2\,v_x+ u_{\mathit{xx}}=0\,,
\end{equation}
сводящейся к
уравнениям Буссинеска заменой $v \to v+\mfrac12\,u_x$. Полный набор
соотношений для вывода главных формул следов содержит определение
величины $\gamma_1^{}=\gamma$ и весь набор интегралов $E_k$:
\begin{equation}\label{ideal1}
\left\{
\begin{array}{rcl}
\gamma &=& v-\mfrac13\,u'+\alpha\,u+\alpha^3,\\
-E_1^{{}^{\mathstrut}}&=& u'-2\,v-\alpha\,u+\alpha^3, \\
-6^{{}^{\mathstrut}}\,E_2&=&
u''-(3\,\alpha^2+u)\,(3\,\alpha^2-2\,u)-9\,\alpha\,E_1\\
36\,E_3&=& u'^2-(3\,\alpha^2+u)\,
(15\,\alpha^4-\mfrac{4^{{\mathstrut}}}{3}\,u^2+\alpha^2\,u-12\,E_2^{}+
18\,\alpha\,E_1)-9\,E_1^2\,.
\end{array}
\right.
\end{equation}
Из этой системы сразу видно, что главные
формулы следов вида $u=S_1(\gamma),\,v=S_2(\gamma)$ не существуют.
Добавляя к \reff{ideal1} производные по $x$ от 1-го и 2-го выражения
в \reff{ideal1}, и рассматривая получившееся как идеал в кольце
$\mathbb{Q}(E_k,\,\alpha,\,\gamma)[u'',\,u',\,v',\,v,\,u,\,\gamma']$
получаем, вычисляя базис Грёбнера \cite{71}, главные формулы следов
c производными $\gamma_x^{}$:
$$
u = \frac13\,
\frac{\gamma_x^2+(3\,\alpha\,\gamma-E_2^{})\,\gamma_x^{} -
2\,(3\,\alpha\,\gamma-E_2)^2}
{-\gamma^2+E_1\,\gamma+E_3}-3\,\alpha^2\,,
$$
$$
v = \frac{\alpha}{3}\,\frac{
3\,\alpha^2\,(5\,\gamma^2+E_1\,\gamma+E_3)-3\,\alpha\,\gamma\,
(\gamma_x^{}+4\,E_2)-\gamma_x^2+E_2\,\gamma+2\,E_2^2}
{-\gamma^2+E_1\,\gamma+E_3} +3\,\gamma-E_1\,.
$$
Используя алгебраическую кривую
$$
\mu^3-(3\,\alpha\,\gamma-E_2)\,\mu-\gamma^2+E_1\,\gamma+E_3=0
$$
и  уравнения Дубровина \reff{dubr}
$$
\gamma_x^{}=-3\,\mu^2+3\,\alpha\,\gamma-E_2
$$
получаем  главные формулы следов вида
$$
u=-3\,\mu-3\,\alpha^2,\qquad
v=6\,\alpha\,\mu+3\,\gamma+2\alpha^3-E_1\,.
$$
В этом примере, разумеется, величины $(\gamma,\,\mu)$
выражаются через эллиптические функции и
соответствующие легко выводимые формулы мы не приводим. Они могут быть
получены и непосредственным интегрированием стационарных уравнений \reff{bsq},
полагая $u,\,v$ зависящими от $x-\alpha\,t$.
Полиномиальный анализ не всегда необходим, но уравнения   \reff{dubr} и вторая координата \reff{mu}
содержат всю информацию о следах и автономной форме уравнений Дубровина. Например равенство \reff{mu} само представляет
одну из последних формул следов.

{\sc Пример 7.} Спектральная задача
$$
\Psi'''+u\,\Psi'+v\,\Psi=\lambda\,\Psi\,, \qquad
u\,\Psi''+(3\,\lambda+v-u'+\alpha)\,\Psi'+\left(\mfrac23\,u''-v'+\mfrac23\,u^2
\right) \Psi=\mu\,\Psi\,,
$$
$$
A=u\,,\qquad B=3\,\lambda+v-u'+\alpha\,,  \qquad \mbox{род\ \
$g=3$}\,.
$$
Факторизуя $R(\lambda;\,x)$
$$
3^{-4}\,R(\lambda;\,x)=\lambda^3+\big (v+\alpha-\mfrac23\,u' \big)
\lambda^2+\mfrac{1}{9}\Big(
 \mfrac23\,u^3+u\,v'+u'^2 +3\,(v+\alpha)^2 -4\,(v+\alpha)\,u'\Big)
\lambda+(\ldots)\,,
$$
получаем тождество следов
$$
v+\alpha-\mfrac23\,u'=-\gamma_1^{}-\gamma_2^{}- \gamma_3^{}\,,
$$
из которого можно считать $v$ найденным, если будет найдено
$u$ (а следовательно и $u'$). Найдем его.
Уравнения \reff{dubr} и вторая координата \reff{mu} имеют вид:
\begin{equation}\label{*}
\gamma_k'=u\,\frac{3\,\mu_k^2+Q(\gk)}
{\displaystyle27\,\prod\limits_{j \ne k}\,(\gk-\gamma_j^{})}\,,
\end{equation}
$$
\mu_k^{}=-\frac{27\,\gamma_k^2+9\,(2\,v-u'+2\,\alpha)\,\gk
+u\,u''+u^3+3\,(\alpha+v)\,(\alpha+v-3\,u')}{3\,u}\,.
$$
Используя последнюю формулу, находим однозначно $u,\,v$:
$$
\frac9u = - \sum\limits_{k=1}^{3}\,\frac{\mu_k^{}} {\displaystyle
\prod\limits_{j \ne k}\,(\gk-\gamma_j^{})}\,, \qquad
v=\frac23\,u'-\alpha-\sum\limits_{k=1}^{3}\,\gk\,.
$$
Кривая и задача обращения Якоби выписывается с очевидностью и мы их не приводим.
После того как они предъявлены, само уравнение  \reff{*}
может выступать как главная формула
следов, несимметричная по парам $(\gamma,\,\mu)$, но с производной $\gamma'$.

{\sc Пример 8.} Спектральная задача (род $g=6$)
$$
\Psi'''+u\,\Psi'=\lambda\,\Psi\,,
$$
$$
-3\,(3\,u\,\lambda-u'''-2\,u\,u')\,\Psi''-
\left(27\,\lambda^2-9\,u'\,\lambda+u^{\sss (\mathrm{IV})}+
3\,u'^2-\mfrac43\,u^3\right) \,\Psi'- 6\,\lambda\,(u''+u^2)\,\Psi
=\mu\,\Psi\,.
$$
Этот случай существенно отличается от предыдущих, так как
\begin{equation}\label{Ak}
A(\gk;\pot)=-9\,u\,\gk+3\,u'''+6\,u\,u'\,,
\end{equation}
а факторизация полинома $R$
$$
\mfrac{-1}{3^{10}}\,R(\lambda;\,x)=\lambda^6-\mfrac23\,u'\,\lambda^5-
\mfrac19\,\big(2\,u\,u''-u'^2+\mfrac23\,u^3+\alpha\big)\,\lambda^4
-\mfrac{2}{81}\,\big(u\,u^{\sss (\mathrm{IV})}
 - \cdots- 2\,\alpha\,u'\big)\,\lambda^3+\cdots
$$
с комбинациями самих уравнений
\reff{dubr}
не дает видимый рецепт извлечения главной формулы следов. Но базис
голоморфных дифференциалов извлекается. В самом деле,
используя очевидные тождества
$$
\sum\limits_{k=1}^{6}\,\frac{\gamma_k^{\nu}} {\displaystyle
\prod\limits_{j \ne k}\,(\gk-\gamma_j^{})}=0\,, \qquad \nu=0 \ldots
4
$$
и линейность \reff{Ak} по $\gk$, сразу получаем
$$
\sum\limits_{k=1}^{6}\,\frac{\gamma_k^{\nu}\,d\gk}
{3\,\mu_k^2+Q(\gk)}=0, \qquad \nu=0,\,1,\,2,\,3\,.
$$
Два недостающих голоморфных дифференциала получаются, как и раньше,
подключением координат $\mu_k^{}$ по формуле \reff{mu}:
$$
\mu_k^{} =3\,\frac {81\,\gamma_k^4-27\,u'\,\gamma_k^3 -
\big(3\,u^{\sss (\mathrm{IV})} +15\,u\,u''+5\,u^3+
6\,\alpha\big)\,\gamma_k^2+\cdots} {3\,u\,\gk-u'''-2\,u\,u'}\,.
$$
В результате получаем оставшуюся часть задачи Якоби
$$
\sum\limits_{k=1}^{6}\,\frac{\mu_k^{}\,d\gk}
{3\,\mu_k^2+Q(\gk)}=0\,,\qquad
\sum\limits_{k=1}^{6}\,\frac{\mu_k^{}\,\gk\,d\gk}
{3\,\mu_k^2+Q(\gk)}=-\frac{1}{27}\,dx\,.
$$

Увеличивать примеры нет смысла и поэтому отметим следующее. Вывод
предложенных формул следов имеет исключительно полиномиальный
характер и может использовать любой из удобных методов
полиномиальной алгебры \cite{71}. Подобная техника с симметрическими
функциями, но уже от двух переменных, будет иметь место в
обобщениях, но формулы приобретут более громоздкий вид. Возникает
вопрос: можно ли предъявить универсальную главную формулу следов или
процедуру ее вывода? Это позволило бы объединить в одну, кажущиеся
разнородными приведенные формулы. По соображениям размерности
очевидно, что формулы типа \reff{trace2} заведомо не будут
справедливы для всех родов. Например замена $4 \to 6$ в
\reff{trace2} невозможна. Сравнение этих же примеров показывает, что
высшие тождества следов, определяющие набор $\gk$ в фундаментальном
полиноме $R$, меняются. Тем не менее положительный ответ на этот
вопрос существует, что будет разобрано отдельно.

\newcommand{\bib}[4]
{\bibitem{#1}  \mbox{\textsc{#2}}\,\,\,{\em #4\/} {#3}}

\newcommand{\ye}[1]{(#1)}

\newcommand{\JMP}[1]{\mbox{Journ.\,\,Math.\,\,Phys. \ye{#1}}}
\newcommand{\PLA}[1]{\mbox{Phys.\,\,Lett.\,\,{\bf A} \ye{#1}}}

\thebibliography{99}


\bib{hitchin}{Атья, М., Хитчин, Н.}{М. (1991), 150 стр.}
{Геометрия и динамика магнитных монополей.}

\bib{14}{Ахиезер, Н.И.}
{В кн. Историко-математические исследования \ye{1978},
вып.\,23,  77--86.}{К спектральной теории уравнения Ламе.}

\bib{15}{Ахиезер, Н.И.}{ДАН СССР \ye{1961}, т.\,141(2), 263--266.}
{Континуальные аналоги ор\-то\-го\-наль\-ных многочленов на системе
интервалов.}

\bib{16}{Белоколос, Е.Д., \ Энольский, В.З.}
{\mbox{Функц. Анализ и его Прил.} \ye{1989}, т.\,23(1), 57--58.}
{Эл\-лип\-ти\-чес\-кие солитоны Вердье и теория редукциии
Вейерштрасса.}

\bib{berkovich}{Беркович, Л.М.}{M. (2002), 464 стр.}
{Факторизация и преобразования дифференциальных уравнений.}

\bib{br4}{Брежнев, Ю.В.}{{\tt http://arXiv/nlin.SI/0504051}.}
{Исторические замечания к теории конечнозонного интегрирования.\\}

\bib{brtheta}{Брежнев, Ю.В.}{Препринт. Калининград (2004).}
{О функциях Якоби и Вейерштрасса.}

\bib{bur}{Буртсев, С.П., Захаров, В.Е., Михайлов, А.В.}
{Теор.\,Мат.\,Физика (1987), т.\,70(3), 323--341.} {Метод обратной
задачи с переменным спектральным параметром.}


\bib{28}{Гельфанд, И.М., \ Дикий, Л.А.}{Успехи Мат. Наук \ye{1975}, т.\,XXX(5), 67--100.}
{Асимптотика ре\-золь\-вен\-ты штурм-лиувиллевских уравнений и алгебра уравнений
Кортевега--де Фриза.}

\bib{58}{Гельфанд, И.М., \ Дикий, Л.А.}{\mbox{Функц. Анализ и его Прил.}
\ye{1976}, т.\,10(4), 13--29.}
{Дробные степени операторов и гамильтоновы системы.}

\bib{59}{Гельфанд, И.М., \ Дикий, Л.А.}{\mbox{Функц. Анализ и его Прил.}
\ye{1977}, т.\,11(2),  11--27.}
{Ре\-золь\-вен\-та и гамильтоновы системы.}

\bib{27}{Гельфанд, И.М., \ Дикий, Л.А.}{\mbox{Функц. Анализ и его Прил.}
\ye{1979}, т.\,13(1), 8--20.}
{Ин\-тег\-ри\-ру\-е\-мые нелинейные урав\-не\-ния и теорема Лиувилля.}

\bib{7}{Дубровин, Б.А.}{\mbox{Функц. Анализ и его Прил.}
\ye{1975}, т.\,9(1), 65--66.}
{Обратная задача теории рассеяния для пе\-ри\-о\-ди\-чес\-ких
конечнозонных потенциалов.}

\bib{8}{Дубровин, Б.A.}{\mbox{Функц. Анализ и его Прил.}
\ye{1975}, т.\,9(3),  41--51.}
{Периодическая задача для уравнения Кортевега--де Фриза в классе
конечнозонных потенциалов.}

\bib{70}{Дубровин, Б.А.}{\mbox{Функц. Анализ и его Прил.} \ye{1977}, т.\,11(4),
28--41.}
{Вполне интегрируемые гамильтоновы сис\-те\-мы,
связанные с матричными операторами, и абелевы многообразия.}

\bib{5}{Дубровин, Б.А.}{Успехи Мат. Наук \ye{1981}, т.\,XXXVI(2),  11--80.}
{Тета-функции и  нелинейные уравнения.}

\bib{2}{Дубровин, Б.А.}{Современные проблемы математики.
Итоги науки и тех\-ни\-ки. М. ВИНИТИ.  \ye{1983}, т.\,23,  33--78.}
{Матричные конечнозонные операторы.}

\bib{29}{Дубровин, Б.А., \ Матвеев, В.Б., \ Новиков, С.П.}
{Успехи Мат. Наук \ye{1976}, т.\,XXXI(1),  55--136.}
{Не\-ли\-ней\-ные уравнения типа КдВ, конечнозонные операторы и
абелевы многообразия.}

\bib{26}{Дубровин, Б.А., \ Новиков, С.П.}
{ЖЭТФ \ye{1974}, т.\,67(6),  2131--2143.}
{Периодический и ус\-лов\-но периодический аналоги многосолитонных решений
уравнения Кортевега де Фриза.}

\bib{39}{Захаров, В.Е., Манаков, С.В., Новиков, С.П., Питаевский, Л.П.}
{Под ред. С.П.\,Новикова. М \ye{1980}, 320 c.} {Теория солитонов.
Метод обратной задачи.}

\bib{47}{Ибрагимов, Н.Х.}{М. \ye{1983}, 280\,с.}
{Группы преобразований в математической физике.}

\bib{57}{Итс, А.Р.}
{Диссертация на соискание уч.\,степени канд. физ.-мат. наук. Л.
\ye{1977}.} {Точное ин\-тег\-ри\-ро\-ва\-ние в римановых
$\theta$-функциях нелинейного уравнения Шрёдингера и
модифицированного уравнения Кортевега--де Фриза.}

\bib{56}{Итс, А.Р.}{Зап.\,Научн.\,Сем.\,ЛОМИ \ye{1984}, т.\,133,  113--125.}
{Теорема Лиувилля и метод об\-рат\-ной задачи.}

\bib{17}{Итс, А.Р., \ Матвеев, В.Б.}{Функц. Анализ и его Прил.
\ye{1975}, т.\,9(1), 69--70.}
{Об операторах Хилла с конечным числом лакун.}

\bib{21}{Итс, А.Р., \ Матвеев, В.Б.}{Теор. Мат. Физика \ye{1975}, т.\,23(1),  51--67.}
{Операторы Шрёдингера с ко\-неч\-но\-зон\-ным спектром и $N$-солитонные решения
уравнения
КдВ.}

\bib{71}{Кокс, Д.,  Литтл, Дж., О'Ши, Д.}{М. \ye{2000}, 688 с.}
{Идеалы, многообразия и алгоритмы.}

\bib{korotkin}{Короткин, Д.А., Матвеев, В.Б.}{Функц. Анализ и его Прил.
(2000), т.\,34(4), 18--34.}{О Тэта-функциональных решениях системы
Шлезингера и уравнения Эрнста.}

\bib{24}{Кричевер, И.М.}{Успехи Мат. Наук \ye{1977}, т.\,XXXII(6), 183--208.}
{Методы алгебраической геометрии в теории нелинейных уравнений.}

\bib{25}{Кричевер, И.М.}{\mbox{Функц. Анализ и его Прил.}
\ye{1978}, т.\,12(3),  20--31.}
{Коммутативные кольца обыкновенных дифференциальных операторов.}

\bib{34}{Кричевер, И.М.}{\mbox{Функц. Анализ и его Прил.}
\ye{1980}, т.\,14(4),  45--54.}
{Эллиптические решения уравнения КП и интегрируемые системы частиц.}

\bib{3}{Kpичевер, И.М.}{Современные проблемы математики.
Итоги науки и тех\-ни\-ки. М. ВИНИТИ. \ye{1983}, т.\,23, 79--136.}
{Нелинейные уравнения и эллиптические кривые.}

\bib{lip}{Липовский, В.Д., Матвеев, В.Б., Смирнов, А.О.}
{Записки научн. сем. ЛОМИ (1986), т.\,150, 70--75.}{О связи между
уравнениями Кадомцева--Петвиашвили и Джонсона.}

\bib{mar}{Марченко, В.А.}{Мат. Сборник \ye{1974}, т.\,95(3), 331--356.}
{Периодическая задача Кортевега--де Фриза.}

\bib{matv}{Матвеев, В.Б.}{Успехи Мат. Наук \ye{1975}, т.\,XXX(6), 201--203.}
{Новая схема интегрирования уравнения Кортевега--де Фриза.}

\bib{78}{Матвеев, В.Б.}{Диссертация на соискание уч. степени
доктора физ.-мат. наук. ЛГУ \ye{1982},  225 стр.}
{Алгебро-геометрические и алгебраические методы интегрирования
нелинейных уравнений типа КдФ.}

\bib{79}{Новиков, В.С.}
{Письма в ЖЭТФ \ye{2000}, т.\,72(3), 223--228.}
{Безотражательные потенциалы акустической спектральной задачи.}

\bib{19}{Новиков, С.П.}{\mbox{Функц. Анализ и его Прил.}
\ye{1974}, т.\,8(3), 54--66.}
{Периодическая задача для уравнения Кортевега--де Фриза. I.}

\bib{61}{Смирнов, А.О.}{Теор. Мат. Физика \ye{1996}, т.\,109(3), 347--356.}
{Об одном классе эллиптических решений уравнения Буссинеска.}

\bib{36}{Смирнов, А.О.}{Мат. Сборник \ye{1999}, т.\,190(5), 139--157.}
{Двухзонные эллиптические ре\-ше\-ния уравнения Буссинеска.}

\bib{b1}{Устинов, Н.В., \ Брежнев, Ю.В.}{Успехи Мат. Наук \ye{2002}, т.\,57(1),
167--168.} {О $\Psi$-функции для конечнозонных потенциалов.}

\bib{80}{Чеботарев, Н.Г.}{М. \ye{1948}, 396\,c.}
{Теория алгебраических функций.}

\bib{yal}{Ялунин, С.В.}{Курсовая работа. Калининград \ye{1997}.}
{О функции Эрмита для спектральной задачи $3$-го порядка.}

\bib{74}{Al'ber, S.I.}{Journ.\,London Math.\,Soc. (2) \ye{1979},
v.\,XIX(3), 467--480.}
{Исследование уравнений Кортевега--де Фриза методом рекуррентных
соотношений.}

\bib{baker}{Baker, H.F.}{Cambridge Univ. Press (1907).}
{An Introductory to the Theory of Multiply Periodic Functions.}

\bib{32}{Baker, H.F.}{Proc.\,Royal Soc.\,London A \ye{1928},
v.\,118, 584--593.}{Note on the foregoing paper,
``Commutative ordinary differential operators'', by J.L.Burchnall
and J.W.Chaundy.}

\bib{38}{Belokolos, E.D., \ Bobenko, A.I., \ Enol'skii, V.Z., \ Its, A.R., \
Matveev V.B.}{Springer--Verlag \ye{1994}, 337\,p.}
{Algebro-Geo\-met\-ric Approach to Nonlinear Integrable Equations.}

\bib{75}{B\l{}aszak, M.}
{Journ.\,Nonlin.\,Math.\,Phys. \ye{2000}, v.\,7(2), 213--243.}
{Degenerate Poisson Pencils on Cur\-ves: New Separability Theory.}

\bib{br}{Brezhnev, Yu.V.}{{\tt http://arXiv/math.CA/0111150}.}
{Uniformization: on the Burnside curve $y^2=x^5-x$.}

\bib{br3}{Brezhnev, Yu.V.}{Phil.\,Trans.\,Royal Soc. Ser A: Math. and
Phys.\,Sciences (in press). {\tt http://arXiv/nlin.SI/0505003}}
{What does integrability of finite-gap/soliton potentials mean?}

\bib{BEL}{Buchstaber, V.M., Enolskii, V.Z., Leykin, D.V.}
{Review in Mathematics and Mathematical Physics. Eds. S.\,Novikov \&
I.\,Krichever. (1997), v.\,10(2), 1--125. London:  Gordon and
Breach.}{Kleinian functions, hyperelliptic Jacobians and
applications.}

\bib{31}{Burchnall, J.L., \ Chaundy, T.W.}{Proc.\,London Math.\,Soc.
\ye{1922}, v.\,22, ser.\,2, n.\,1435, 420--440.}
{Commutative ordinary differential operators.}


\bib{84}{Caudrey, P.J.}{\PLA{1980}, v.\,79, 264--268.}
{The inverse problems for the third order equation
$u_{\mathit{xxx}}+q(x)\,u_x+r(x)\,u=-i\,\zeta^3\,u$.}

\bib{66}{Chazy, J.}{Acta Mathematica \ye{1911}, v.\,34, 317--385.}
{Sur les \'equations diff\'erentielles du troisi\'eme
ordre et d'ordre sup\'erieur dont l'int\'egrale g\'en\'erale a ses
points critiques fixes.}

\bib{53}{Deift, P., \ Tomei, C., \ Trubowitz, E.}{Comm.\,Pure and Appl.\,Math.
\ye{1982}, v.\,35, 567--628.}
{In\-ver\-se Scattering and the  Boussinesq Equation.}

\bib{dik1}{Dickey, L.A.}{Comm.\,Math.\,Phys. \ye{1981}, v.\,82(3),
345--360, 361--375.} {Integrable Nonlinear Equations and Liouville's
Theorem, I, II\/.}


\bib{dik3}{Dickey, L.A.}{Comm.\,Math.\,Phys. \ye{1983}, v.\,88(1), 27--42.}
{Hamiltonian Structures and Lax Equations Generated by Matrix Differential
Operators with polynomial Dependence on a Parameter\/.}

\bib{dik4}{Dickey, L.A.}
{Advance Series in Math.\,\,Physics 12. World Scientific \ye{1991},
310 pp.}{Solitons equations and hamiltonian systems.}

\bib{44}{Dickson, R., \ Gesztesy, F., \ Unterkofler, K.}
{Rev.\,\,Math.\,Phys. \ye{1999}, v.\,11(7), 823--879.}
{Al\-geb\-ro-ge\-o\-met\-ric solutions of the Boussinesq hierarchy.}

\bib{45}{Dickson, R., \ Gesztesy, F., \ Unterkofler, K.}
{Math.\,Nachr. \ye{1999},
v.\,198, 51--108.}{A new approach to the Boussinesq hierarchy.}

\bib{dmitr}{Dmitrieva, L.A.}{\PLA{1993}, v.\,182, 65--70.}
{Finite-gap solutions of the Harry Dym equation.}

\bib{49_1}{Drach, J.}{Compt. Rend. Acad. Sci. \ye{1919}, t.\,168, 47--50.}
{D\'etermination des cas de r\'eduction de l'\'equation diff\'erentielle
$\frac{d^2y}{dx^2}=[\varphi(x)+h]\,y$.}

\bib{49_2}{Drach, J.}{Compt. Rend. Acad. Sci. \ye{1919}, t.\,168, 337--340.}
{Sur l'integration par quadratures de l'equation
$\frac{d^2y}{dx^2}=[\varphi(x)+h]\,y$.}



\bib{43}{Gesztesy, F., \ Ratnaseelan, R.}{Rev.\,\,Math.\,Phys. \ye{1998},
v.\,10, 345--391.}{An alternative approach to algebro--geometric solutions of
the AKNS hierarchy.}

\bib{46}{Gesztesy, F., \ Holden, H.}
{Math. Scand. \ye{2001}, v.\,90, 1--36.}
{Dubrovin equations and integrable systems on hyperelliptic curves.}

\bib{gesztesy}{Gesztesy, F., Holden, H.}{Cambridge (2003).}{Soliton equations and their algebro-geometric solutions.}


\bib{68}{Hereman, W., \ G\"oktas \"U.}
{{\tt http://xxx.lanl.gov: solv-int/9904022.}}
{Integrability Tests for Nonlinear Evolution Equations.}

\bib{kapl}{Kaplansky, I.}{Paris: Hermann \ye{1957}, 62 pp.}
{An introduction to differential algebra.}

\bib{64}{Kaup, D.J.}{Stud.\,Appl.\,Math. \ye{1980}, v.\,62, 189--216.}
{On the inverse scattering problem for cubic eigenvalue problems
of the class $\psi_{\mathit{xxx}}^{}+6\,Q\,\psi_x^{}+6\,R\,\psi=\lambda\,\psi$.}

\bib{kolchin}{Kolchin, E.R.}{Annals of Math. \ye{1948}, v.\,49, 1--42.}
{Algebraic matrix groups and the Picard--Vessiot theory of
homogeneous linear ordinary differential equations.}

\bib{kov}{Kovacic, J.J.}{Journ. Symbolic Comput. \ye{1986}, v.\,2(1), 3--43.}
{An algorithm for solving second order linear
homogeneous differential equations.}

\bib{kuz}{Kuznetsov, V.B.,  Nijhoff, F.W., Sklyanin, E.K.}
{Comm.\,\,Math.\,Phys. (1997), v.\,189(3), 855--877.}
{Separation of variables for the Ruisenaars system.}

\bib{55}{Lax, P.D.}{Comm.\,Pure and Appl.\,Math. \ye{1975}, v.\,28, 141--188.}
{Periodic Solutions of the KDV-equation.}

\bib{4}{Matveev, V.B.}{Wroc\l{}aw University \ye{1976}. Preprint  n.\,373,
1--98.}
{Abelian functions and solitons.}

\bib{37}{McKean, H.P., \ van Moerbeke, P.}
{Inventions in Mathematics \ye{1975},
v.\,30(3), 217--274.}{The spectrum of Hill's equation.}

\bib{54}{McKean, H.P.}
{Comm.\,Pure and Appl.\,Math. \ye{1981}, v.\,34, 599--691.}
{Boussinesq's Equation on the Circle.}

\bib{22}{Musette, M., \ Conte, R.}{\JMP{1998}, v.\,39(10), 5617--5630.}
{B\"acklund transformation of partial differential
equations from the Painlev\'e--Gambier classification. I. Kaup--Kupershmidt
equation.}

\bib{ispanec}{Morales-Ruiz, J.J.}
{Progress in Mathematics, {\bf 179}. Basel: Birkh\"auser Verlag (1999).}{Differential Galois theory and
non-integrability of Hamiltonian systems.}

\bib{76}{Musette, M., \ Verhoeven, C.}{Physica D \ye{2000}, v.\,144(1--2),
211--220.}{Nonlinear superposition formula for the Kaup--Kupershmidt
partial differential equation.}

\bib{prym}{Prym, F.}{Berlin: Mayer \& M\"uller (1885).}
{Neue Theorie der ultraelliptischen functionen.}

\bib{singerput}{van der Put, M., \ Singer, M. F.}
{Berlin: Springer--Verlag (2003).}
{Galois theory of linear differential equations.}

\bib{65}{Sawada, K., \ Kotera, T.}{Progr.\,Theor.\,Phys. \ye{1974}, v.\,51(5),
1355--1367.}{A Method for Finding N-Soliton Solutions of K.d.V. Equation
and K.d.V.-Like Equation.}

\bib{singer}{Singer, M.F.}{Amer.\,\,Journ.\,\,Math. \ye{1981}, v.\,103,
661--682.}{Liouvillian solutions of $n$th order homogeneous
linear differential equations.}

\bib{33}{Smirnov, A.O.}{Acta Appl.\,Math. \ye{1994}, v.\,36, 125--166.}
{Finite-gap Elliptic Solutions of the KdV Equation.}

\bib{smirnov}{Smirnov,  A.O.}
{The Kowalevski property (Leeds, 2000), CRM Proc. Lecture Notes 32,
287--305. Amer.\,Math.\,Soc., Providence, RI, \ye{2002}.} {Elliptic
solitons and Heun's equation.}

\bib{tah}{Takhtajan, L.A.}{Теор. Мат. Физика (1992), т.\,93(2),
330--341.}{A simple example of modular forms as tau-functions for
integrable equations.}

\end{document}